\documentclass[11pt,twoside]{article}
\pdfoutput=1     

\usepackage{pslatex}    
\usepackage{amsfonts}
\usepackage{amsmath}
\usepackage{amsthm}
\usepackage{amscd}
\usepackage{cite}
\usepackage{eucal}
\usepackage{graphicx}
\usepackage{graphics, color}
\usepackage{a4}
\usepackage{booktabs}
\usepackage{microtype}
\usepackage{dsfont}

\def\beq{\begin{equation}}
\def\eeq{\end{equation}}
\def\bea{\begin{eqnarray}}
\def\eea{\end{eqnarray}}

\def\beann{\begin{eqnarray*}}
\def\eeann{\end{eqnarray*}}
\def\lhs{left hand side}
\def\rhs{right hand side}
\def\wtrho#1{\widetilde\rho_{#1}}
\def\ntrho#1{\rho_{#1}}

\let\a=\alpha \let\be=\beta  
   
\let\eps=\epsilon
  \let\la=\lambda \let\m=\mu
 \let\x=\xi \let\p=\pi \let\r=\rho \let\s=\sigma
 
 \let\Ph=\phi  \let\Ps=\Psi
  
\let\La=\Lambda \let\G=\Gamma 
\let\vphi=\varphi
\def\xx{\tilde{\xi}}

\let\qd=\quad  

\def\epp{\, .}
\def\epc{\, ,}

\theoremstyle{plain}

\newtheorem*{corollary*}{Corollary}

\newtheorem*{conjecture*}{Conjecture}

\theoremstyle{definition}

\def\2{\frac{1}{2}} \def\4{\frac{1}{4}}

\def\6{\partial}

\def\+{\dagger}

\def\<{\langle} \def\>{\rangle}

\def\CO{{\cal O}}

\def\i{{\rm i}}

\def\rd{{\rm d}}
\def\re{{\rm e}}

\def\ctg{\, {\rm ctg}\,}

\renewcommand{\cot}{\ctg}

\DeclareMathOperator{\tr}{tr}

\DeclareMathOperator{\End}{End}

\DeclareMathOperator{\detq}{{\rm det}_q}

\def\Bf{\mathfrak{B}}
\def\Bfq{\overline{\mathfrak{B}}}
\def\fb{\mathfrak{b}}
\def\fbq{\overline{\mathfrak{b}}}

\def\jg2{K}

\renewcommand{\appendix}{%
   \renewcommand{\section}{
        \secdef\Appendix\sAppendix}%
   \setcounter{section}{0}%
   \renewcommand{\thesection}{\Alph{section}}%
   \renewcommand{\theequation}{\thesection.\arabic{equation}}%
}

\newcommand{\Appendix}[2][?]{%
     \refstepcounter{section}%
     \setcounter{equation}{0}%
     \addcontentsline{toc}{appendix}%
          {\protect\numberline{\appendixname~\thesection} #1}%
     \vspace{\baselineskip}%
     {\noindent\large\bfseries\appendixname\ \thesection: #2\par}%
     \sectionmark{#1}\vspace{\baselineskip}}

\newcommand{\sAppendix}[1]{%
     {\noindent\large\bfseries\appendixname\:: #1\par}%
     \sectionmark{#1}\vspace{\baselineskip}}

\renewcommand{\tilde}{\widetilde}
\begin{document}

\thispagestyle{empty}

\begin{center}

{\Large {\bf Correlation functions of the integrable isotropic
spin-1 chain: algebraic expressions for arbitrary temperature
\\}}

\vspace{7mm}

{\large
Andreas Kl\"{u}mper\footnote[1]{e-mail: kluemper@uni-wuppertal.de},
Dominic Nawrath\footnote[2]{e-mail: dnawrath@uni-wuppertal.de}}%
\\[1ex]
Fachbereich C -- Physik, Bergische Universit\"at Wuppertal,\\
42097 Wuppertal, Germany\\[2.5ex]
{\large Junji Suzuki\footnote[3]{e-mail: sjsuzuk@ipc.shizuoka.ac.jp}}%
\\[.5ex]
Department of Physics, Faculty of Science, Shizuoka University,\\
Ohya 836, Suruga, Shizuoka, Japan

\vspace{40mm}

{\large {\bf Abstract}}

\end{center}

\begin{list}{}{\addtolength{\rightmargin}{9mm}
               \addtolength{\topsep}{-5mm}}
\item
We derive algebraic formulas for the density matrices of finite segments
of the integrable $su(2)$ iso\-tropic spin-1 chain in the thermodynamic limit.
We give explicit results for the 2 and 3 site cases for arbitrary temperature
$T$ and zero field. In the zero temperature limit the correlation functions
are given in elementary form in terms of Riemann's zeta function at even
integer arguments.
\\[2ex]
\end{list}

\clearpage

\section{Introduction}
In recent years an abundance of knowledge on the correlation functions of
integrable lattice systems has been obtained. Most of the development took
place for the static correlations of systems with spin-1/2 representations of
the algebra $su(2)$ including various anisotropic and inhomogeneous
deformations.

The developments started with the discovery of multiple integral
representations for the density matrix of a finite chain segment
\cite{JMMN92,JiMi96,KMT99b,GKS05} and the observation \cite{BoKo01} of the
factorization of these integrals into sums over products of single
integrals. With \cite{BGKS06,DGHK07} it became apparent that existence and
factorization of multiple integral formulas are not tied to special properties
of ground-states of infinite systems that often have higher symmetries. This
holds even for finite temperature in the thermodynamic limit as well as for the ground
states of finite chains motivating the research
\cite{BJMST06b,BJMST08a,JMS08,BoGo09}. Multiple integral formulations led to
ab initio calculations of the asymptotics of ground state correlators of the
XXZ chain\cite{KKMST09}. Results for the closely related scalar Bose gas and
for the Sine-Gordon model were obtained in
\cite{KKMST07,SBGK07a,SGK07,JMS09pp}.

Meanwhile the `hidden Grassmann structure' was identified \cite{BJMST06b,
  BJMST08a} and made it possible to prove the complete factorization of
general static correlation functions in terms of elementary nearest-neighbour
functions under very general conditions \cite{JMS08,BoGo09}.

In our view the most interesting question is if and how the investigations can
be extended to models based on representations of higher rank algebras or to
those based on higher level representations of $su(2)$. 
We believe that the latter goal is in closer reach, but still very
challenging.
In particular, on the basis of \cite{JMS08} we expect algebraically transparent
and physically relevant expressions for higher-spin generalizations
constructed by means of the fusion procedure \cite{KuSk82b,KRS81}. 

Some older results of multiple integral type exist for the ground state
and were derived in the $q$-vertex operator approach in \cite{Idzumi94,BoWe94}
and in the combinatorial application of the algebraic Bethe ansatz in
\cite{Kitanine01} for the isotropic case which was generalized to the zero
field XXZ-case in \cite{DeMa10}. The generalization of \cite{Kitanine01} to
finite temperature and field was undertaken in \cite{GSS10}. Unfortunately, the
resulting formulae are rather intricate for applications. In the present
publication we undertake an effort to directly construct algebraic expressions
for the finite temperature density matrices containing all information on
correlation functions restricted to finite segments of infinitely long chains.
Finite temperatures are treated by suitable `lattice path integral'
formulations, i.e. mapping the 1d quantum chain to a suitable 2d classical
system. Physical quantities are obtained by a combination of algebraic
techniques ($T$- and $Y$-systems of transfer matrices) and analytical means
(functional equations).

A basic notion in the `lattice path integral' approach to finite temperatures
is the quantum transfer matrix (QTM) \cite{Suzuki85} acting in chain direction
and being the `quantum analogue' of quantum chains to the `classical transfer
matrix' of the classical Ising spin chain. The QTM satisfies the so-called
$Y$-system from which by algebraic and analytic means the well-known
thermodynamical Bethe ansatz (TBA) equations can be derived
\cite{JKS98a,KP92}. Another way of analysis makes use of a finite number of
auxiliary functions which satisfy a closed set of functional equations. This
approach was developed for the spin-1/2 case of XXX, XXZ and XYZ
\cite{Kluemper93} and is probably the least canonical formulation, but highly
efficient especially in applications. The generalization to the higher spin
case was not clear before the work \cite{Suzuki99}. As we shall see below the
auxiliary functions introduced in \cite{Suzuki99} are indeed most useful also
for the explicit computation of correlation function of higher spin systems.

In this work we derive algebraic expressions for the correlation functions
of the integrable $su(2)$ invariant spin-1 chain at arbitrary temperature, but
zero field. We will be explicit for the 2-site case and give results for 3
sites. As the reader will see, the generalization to more sites and higher
spin systems is viable.

The paper is organized as follows. In section 2 we recall the construction of
the Hamiltonian, the fusion procedure and fundamentals of thermodynamics. In
section 3 we introduce auxiliary density matrices and computational `tricks',
especially functional equations and the algebraic Bethe ansatz for a QTM with
modifications yielding a `generating function' for the fundamental
nearest-neighbour correlator in terms of which any other static correlation
function factorizes. The fundamental nearest-neighbour correlator is
characterized in section 4 in terms of useful integral equations. Explicit
results in the low- and high-temperature limits are given.  In section 5 we
present concrete results for 2 sites. Section 6 is devoted to the complete
computation of the 3 site density matrix. A conclusion is given which is
followed by appendices containing details of our construction.

\section{Fundamental thermodynamical and integrable structures}
\subsection{Hamiltonian}
The Hamiltonian of the integrable isotropic spin-1 chain on a lattice
of $L$ sites is
\begin{equation} \label{ham}
     H = \frac{J}{4} \sum_{n = 1}^L
           \bigl[ \vec S_{n-1}\cdot \vec S_n - (\vec S_{n-1}\cdot\vec  S_n)^2
                  \bigr] \epc
\end{equation}
with for instance periodic
boundary conditions $\vec S_{L+1} = \vec S_1$.
The spin components $S_n^\a$ act locally as standard spin-1 operators, and
antiferromagnetic exchange, $J > 0$, is assumed throughout the paper.

The Hamiltonian (\ref{ham}) was first obtained in a more general
anisotropic form in \cite{ZaFa80}. Shortly later it was constructed
by means of the fusion procedure \cite{KuSk82b,KRS81}. The ground state
and the elementary excitations were studied in \cite{Takhtajan82},
and an algebraic Bethe ansatz and the thermodynamics within the TBA
approach were obtained in \cite{Babujian82}.

\subsection{Integrable structure}
The model can be constructed by means of the fusion procedure
\cite{KRS81}, starting from the fundamental spin-$\2$ $R$-matrix 
\begin{equation}
     R^{[1,1]} (\la) = \begin{pmatrix} 1 &&& \\ & b(\la) & c(\la) & \\
                              & c(\la) & b(\la) & \\ &&& 1
              \end{pmatrix} \epc \qd
	      b(\la) = \frac{\la}{\la + 2 \i} \epc \qd
	      c(\la) = \frac{2 \i}{\la + 2 \i} \epc
\end{equation}
which we think of as an element of $\End ({\mathbb C}^2 \otimes
{\mathbb C}^2)$ (superscripts 1 indicate the spin-1/2 aka level 1 representations). 
It satisfies the Yang-Baxter equation
\begin{equation} \label{ybe}
     R^{[1,1]}_{12} (\la - \m) R^{[1,1]}_{13} (\la) R^{[1,1]}_{23} (\m)
      = R^{[1,1]}_{23} (\m) R^{[1,1]}_{13} (\la) R^{[1,1]}_{12} (\la - \m)
        \epp
\end{equation}
As usual the $R^{[1,1]}_{jk}$ in this equation act on the $j$th and $k$th
factor of the triple tensor product ${\mathbb C}^2 \otimes {\mathbb C}^2
\otimes {\mathbb C}^2$ as $R^{[1,1]}$ and on the remaining factor
trivially. $R^{[1,1]}$ is normalized in such a way that
\begin{equation}
     R^{[1,1]} (0) = P^{[1]} \epc
\end{equation}
where $P^{[1]}$ is the transposition of the two factors in ${\mathbb C}^2
\otimes {\mathbb C}^2$. We say that $R^{[1,1]}$ is regular. At the same
time $\check R^{[1,1]} = P^{[1]} R^{[1,1]}$ satisfies the unitarity
condition
\begin{equation} \label{uni1}
     \check R^{[1,1]} (\la) \, \check R^{[1,1]} (-\la) = 
\mathds{1}_4\epc
\end{equation} 
with $\mathds{1}_n$ denoting the $n \times n$ unit matrix.

A further property of $R^{[1,1]}$, which is at the heart of the fusion
procedure, is its degeneracy at two special points,
\begin{equation} \label{degen}
     \lim_{\la \rightarrow \pm 2 \i} \frac{R^{[1,1]} (\la)}{2 b(\la)}
        = P^\pm \epc 
\end{equation}
where $P^\pm$ are the orthogonal projectors onto the singlet and triplet
subspaces $V^{(s)}, V^{(t)} \subset {\mathbb C}^2 \otimes {\mathbb C}^2$.
Due to (\ref{ybe}) and (\ref{degen}) we have the important relation
\begin{equation} \label{tripinvar}
     P_{23}^- R^{[1,1]}_{13} (\la) R^{[1,1]}_{12} (\la + 2 \i) P_{23}^+ = 0
        \epc
\end{equation}
meaning that $R^{[1,1]}_{13} (\la) R^{[1,1]}_{12} (\la + 2 \i)$ leaves
${\mathbb C}^2 \otimes V^{(t)}$ invariant.

Be $S: {\mathbb C}^2 \otimes {\mathbb C}^2 \rightarrow {\mathbb C}^3$
the projector onto the triplet space
we write the fused $R$-matrices as
\begin{subequations}
\label{fusedr}
\begin{align}
     R^{[1,2]} (\la) & =
        S_{23}\, R_{13}^{[1,1]} (\la)
	         R_{12}^{[1,1]} (\la + 2\i)\, S_{23}^t \epc \\[1ex]
     R^{[2,1]} (\la) & =
        S_{12}\, R_{13}^{[1,1]} (\la - 2\i)
	         R_{23}^{[1,1]}\, (\la) S_{12}^t \epc \\[1ex]
     R^{[2,2]} (\la) & = S_{12} S_{34} R^{[1,1]}_{14} (\la - 2 \i)
                 R^{[1,1]}_{13} (\la) R^{[1,1]}_{24} (\la)
		 R^{[1,1]}_{23} (\la + 2 \i) S_{34}^t S_{12}^t
\end{align}
\end{subequations}
acting on ${\mathbb C}^2 \otimes {\mathbb C}^3$, ${\mathbb C}^3 \otimes
{\mathbb C}^2$, or ${\mathbb C}^3 \otimes {\mathbb C}^3$, respectively.
Superscrips 1 and 2 refer to level 1 and level 2 (spin-1/2 and spin-1) representations.
Combining the Yang-Baxter equation (\ref{ybe}) and equations
(\ref{tripinvar}) it is easy to see that
\begin{equation} \label{ybes}
     R^{[2s_1,2s_2]}_{12} (\la - \m) R^{[2s_1,2s_3]}_{13} (\la)
        R^{[2s_2,2s_3]}_{23} (\m)
      = R^{[2s_2,2s_3]}_{23} (\m) R^{[2s_1,2s_3]}_{13} (\la)
        R^{[2s_1,2s_2]}_{12} (\la - \m) \epc
\end{equation}
where $s_j = \2, 1$ for $j = 1, 2, 3$.

In particular, $R^{[2,2]}$ is a solution of the Yang-Baxter equation.
With $P^{[2]}$ denoting the transposition on ${\mathbb C}^3 \otimes
{\mathbb C}^3$ and $\check R^{[2,2]} = P^{[2]} R^{[2,2]}$ we have the standard
initial condition and unitarity in the following form
\begin{subequations}
\label{regun}
\begin{align} \label{regular}
     & R^{[2,2]} (0) = P^{[2]} \epc \qd
       \\[1ex] \label{unitary}
     & \check R^{[2,2]} (\la)\, \check R^{[2,2]} (-\la) = \mathds{1}_9 \epp
\end{align}
\end{subequations}
It follows with (\ref{regular}) that $R^{[2,2]}$ generates
the Hamiltonian (\ref{ham}),
\begin{equation} \label{locham}
     H = \i J \sum_{n = 1}^L h_{n-1, n} \epc \qd h_{n-1, n} =
         \6_\la \check R^{[2,2]}_{n-1, n} (\la) \bigr|_{\la = 0} \epp
\end{equation}
\subsection{Physical density matrix}
In \cite{GKS04a} we have set up a formalism which enables us to calculate
thermal correlation functions in integrable models with $R$-matrices
fulfilling (\ref{regular}). It is based on the so-called quantum transfer
matrix \cite{Suzuki85} and its associated monodromy matrix which are
directly related to the statistical operator.

The Hamiltonian (\ref{ham}) preserves the total spin $S^\a = \sum_{j = 1}^L
S_j^\a$. For instance, the magnetization in $z$-direction is a thermodynamic
quantity, and the statistical operator
\begin{equation}
     \r_L (T, h) = \re^{ - \beta{H - 2 h S^z}}
\end{equation}
describes the spin chain (\ref{ham}) in thermal equilibrium at temperature
$T (=1/\beta)$ and magnetic field~$h$. 

The free energy per lattice site $f(T,h)$ determines the macroscopic
thermodynamics of the model. For the general integrable spin-$S$ Heisenberg
chain the free energy was calculated in \cite{Suzuki99}.

The free energy is related to the partition function of the
statistical operator $Z:=\tr \r_L$. The normalized statistical operator
\begin{equation}
     D_L:=\frac 1Z \r_L,
\end{equation}
is known as the density operator of the total system.

It is convenient to define the reduced density matrix of a finite chain
segment $[1,m]$ by tracing out all other degrees of freedom
\begin{equation} \label{defdensmat}
     D_{[1,m]} := \tr_{\{m + 1, \dots, L\}}\, D_L \epp
\end{equation}
The reduced density matrix is well-defined even in the thermodynamic limit
$L\to\infty$.  With $D_{[1,m]}$ we can calculate the expectation value of any
local operator that acts trivially outside the finite segment $[1,m]$.

For any integrable model, whose $R$-matrix does not only satisfy the
Yang-Baxter equation, but also the regularity and unitarity conditions
(\ref{regun}), we can approximate the statistical operator $\r_L (T, h)$
of the $L$-site Hamiltonian using the monodromy matrix of an
appropriately defined vertex model with $L$ vertical lines ($1,
\dots, L$) and $N$ alternating horizontal lines ($\overline 1, \dots,
\overline N$ with $N$ even). This fact was exploited many times in the
calculation of the bulk thermodynamic properties of integrable quantum
chains, in particular, in case of the higher-spin integrable Heisenberg
chains \cite{Suzuki99}. In \cite{GKS04a} it was noticed that the same
formalism is also useful for the calculation of thermal correlation
functions. Following the general prescription in \cite{GKS04a} we define
\begin{multline}
     T_j^{[2]} (\la) = \re^{2 h S_j^z/T}
        R_{j, \overline N}^{[2,2]} (\la +\i u)
        R_{\overline{N-1}, j}^{[2,2] \: t_1} (\i u - \la) \dots \\ \dots
        R_{j, \overline 2}^{[2,2]} (\la +\i u)
        R_{\overline 1, j}^{[2,2] \: t_1} (\i u - \la) \epc
\end{multline}
where $u:=-J\be/N$ and $t_1$ indicates transposition with respect to the first
space in a tensor product. 
This monodromy matrix is constructed in such a way that
(see \cite{GKS04a})
\begin{equation} \label{piovertwo}
      \tr_{\bar 1 \dots \overline N} \Bigl\{
         T^{[2]}_{1} (0) \dots T^{[2]}_L (0) \Bigr\}
         = \biggl[1 - \frac{2}{NT} \sum_{n = 1}^L
                      \bigr(J h_{n-1,n} - 2 h S_n^z \bigr)
		      + \CO \Bigl( \frac{1}{N^2} \Bigr)
                      \biggr]^\frac{N}{2} \mspace{-9.mu} .
\end{equation}
Hence, the statistical operator $\r_L$ can be approximated by
$\r_{N, L} (T, h) := $\\ $\tr_{\bar 1 \dots \overline N} \Bigl\{
         T^{[2]}_{1} (0) \dots T^{[2]}_L (0) \Bigr\}$.
In the so-called Trotter limit of $N\to\infty$ we have $\r_{N,L}\to\r_L$.

The transfer matrix
\begin{equation}
     t^{[2]} (\la ) = \tr_j T^{[2]}_j (\la )
\end{equation}
is commonly called the quantum transfer matrix. We shall recall below
how it can be diagonalized by means of the algebraic Bethe ansatz
\cite{Suzuki99}. Quite generally it has the remarkable property that
the eigenvalue $\La^{[2]} (0)$ of largest modulus of $t^{[2]} (0)$ (we
call it the dominant eigenvalue) is real and non-degenerate and is
separated by the rest of the spectrum by a gap \cite{Suzuki85,SuIn87,SAW}.
It can further be shown that
\begin{equation} \label{free}
     f (T,h) = - T \lim_{N \rightarrow \infty} \ln \La^{[2]} (0) \epp
\end{equation}
Thus, the dominant eigenvalue alone determines the bulk thermodynamic
properties of the spin chain.

Owing to the fact that $R^{[2,2]}$ satisfies the Yang-Baxter equation
the transfer matrices $t^{[2]} (\la)$ form a commuting family,
\begin{equation}
     [t^{[2]} (\la), t^{[2]} (\m)] = 0 \epp
\end{equation}
It follows that the eigenvectors of $t^{[2]} (\la)$ do not depend on
$\la$. Let $|\Ps\>$ denote a normalized eigenvector belonging to the
dominant eigenvalue $\La^{[2]} (0)$. We shall call it the dominant
eigenvector.  It is unique up to normalization and is an eigenvector of
$t^{[2]} (\la)$ with eigenvalue $\La^{[2]} (\la) = \<\Ps| t^{[2]}
(\la)|\Ps\>$. In \cite{GKS04a} it was pointed out that such an eigenvector
determines all static correlation functions at temperature $T$ and magnetic
field $h$. In particular, it determines the density matrix (\ref{defdensmat})
of any finite segment $[1,m]$,
\begin{equation} \label{d2hom}
     D_{[1,m]} (T, h) = \lim_{N \rightarrow \infty}
        \frac{\<\Ps| T^{[2]} (0) \otimes \dots \otimes
                       T^{[2]} (0) |\Ps\>}
             {\bigl( \La^{[2]} (0)\bigr)^m} \epp
\end{equation}

For technical reasons it is better to consider a slightly more general
expression than the one under the limit, by allowing for mutually
distinct spectral parameters $\x_j$, $j = 1, \dots, m$, instead of zero.
Setting $\x = (\x_1, \dots, \x_m)$ we define
\begin{equation} \label{d2inhom}
     D^{[2]} (\x_1,...,\x_m) = \frac{\<\Ps| T^{[2]} (\x_1) \otimes \dots
                          \otimes T^{[2]} (\x_m) |\Ps\>}
                         {\La^{[2]} (\x_1) \dots
                          \La^{[2]} (\x_m)} \epc
\end{equation}
the inhomogeneous density matrix at finite Trotter number. Then
\begin{equation} \label{d2lim}
     D_{[1,m]} (T, h) = \lim_{N \rightarrow \infty} \:
                      \lim_{\x_1, \dots, \x_m \rightarrow 0}
                      D^{[2]} (\x) \epp
\end{equation}

\section{Modified lattice model and auxiliary density matrices}
The physically interesting correlation functions are spin-1 correlators,
however, the most fundamental objects are spin-1/2 correlators resp.~spin-1/2
density matrices
\begin{equation} \label{d2inhom1}
     D^{[1]} (\xx_1,...,\xx_n) = \frac{\<\Ps| T^{[1]} (\xx_1) \otimes \dots
                          \otimes T^{[1]} (\xx_n) |\Ps\>}
                         {\La^{[1]} (\xx_1) \dots
                          \La^{[1]} (\xx_n)} \epp
\end{equation}
Both types of density matrices are properly normalized, i.e. the total traces
evaluate to 1, because of the normalization terms in the denominators. Taking
traces of $D^{[j]}$ turns the monodromy matrices in the numerators
$T^{[j]}(\xi)$ into transfer matrices $t^{[j]}(\xi)$ which evaluate to the
eigenvalue $\La^{[j]}(\xi)$ when acting on the eigenstate $|\Ps\>$. 

The density matrix $D^{[2]} (\x_1,...,\x_m)$ is obtained from $D^{[1]}
(\xx_1,...,\xx_n)$ by fusion in the case $n=2m$ and symmetrizations over
neighbour pairs of quantum spaces carrying spectral parameters
$\xx_{2k-1}=\xi_k-\i$, $\xx_{2k}=\xi_k+\i$. This procedure yields the
unnormalized $D^{[2]} (\x_1,...,\x_m)$. The normalization factor is given by
the ratio of denominators occuring in (\ref{d2inhom1}), (\ref{d2inhom})
\begin{equation} \label{normalfact}
\prod_{k=1}^m
\frac{\La^{[1]} (\xi_k-\i)\La^{[1]} (\xi_k+\i)}{\La^{[2]} (\xi_k)}.
\end{equation}
In this paper we follow the strategy of (discrete) functional equations for
finite temperature \cite{AuKl12} to calculate the spin-1 correlators.
This approach is actually formulated and viable for finite Trotter number $N$.
It yields $N$ many equations for the density matrix $D^{[2]} (\x_1,...,\x_m)$
as function of {\em one} of its spectral parameter arguments, let us say the
last one $\x_m$. Together with the asymptotic condition
$\lim_{\x_m\to\infty}D^{[2]} (\x_1,...,\x_{m-1},\x_m)=D^{[2]}
(\x_1,...,\x_{m-1})\otimes \mathds{1}$ (for zero field) it fixes the dependence of $D^{[2]}
(\x_1,...,\x_m)$ on $\x_m$ if the data $D^{[2]}(\x_1,...,\x_{m-1})$ are given
and suitable analyticity conditions hold. The analyticity
conditions in the spin-1/2 case \cite{AuKl12} are simple: The entries
of the density matrix for a system with Trotter number $N$ are multivariate
polynomials of degree $N$ divided by the known multivariate
polynomial $\La(\x_1)...\La(\x_m)$. Hence, the $N+1$ many equations (linear
independence was proven in \cite{AuKl12}) fix the $N+1$ many coefficients.

For our case at hand two different approaches are in principle conceivable:
(i) the spin-1 object $D^{[2]} (\x_1,...,\x_m)$ is tackled directly, or (ii) 
the spin-1/2 object $D^{[1]} (\xx_1,...,\xx_n)$ is treated first and then 
fusion is applied.
Unfortunately, both approaches present problems: (i) the object $D^{[2]}
(\x_1,...,\x_m)$ consists of multivariate polynomials of degree $2N$ (!);
(ii) for $D^{[1]} (\xx_1,...,\xx_n)$ which consists of multivariate
polynomials of degree $N$ the functional equations can not be derived.
In both cases though, the problems can be overcome.

\subsection{Functional properties of $D^{[2]} (\x_1,...,\x_m)$}
In this paragraph we review the basic constructions for setting up (discrete)
functional equations for reduced density matrices of integrable vertex models
on semi-infinite ($\infty\times N$) square lattices. It is possible, and for
later purposes convenient, to start
with a system with $N$ many different, but fixed spectral parameters $u_j$ on
the $N$ many horizontal lines. The (unnormalized) density matrix on a finite
sequence of sites has matrix elements that are partition functions of
`sliced lattices'. For notational simplicity we write $D(\x_1,...,\x_m)$
for $D^{[2]} (\x_1,...,\x_m)$ wherever possible.
\begin{figure}[h]
\begin{center}
\includegraphics*[width=0.7\textwidth]{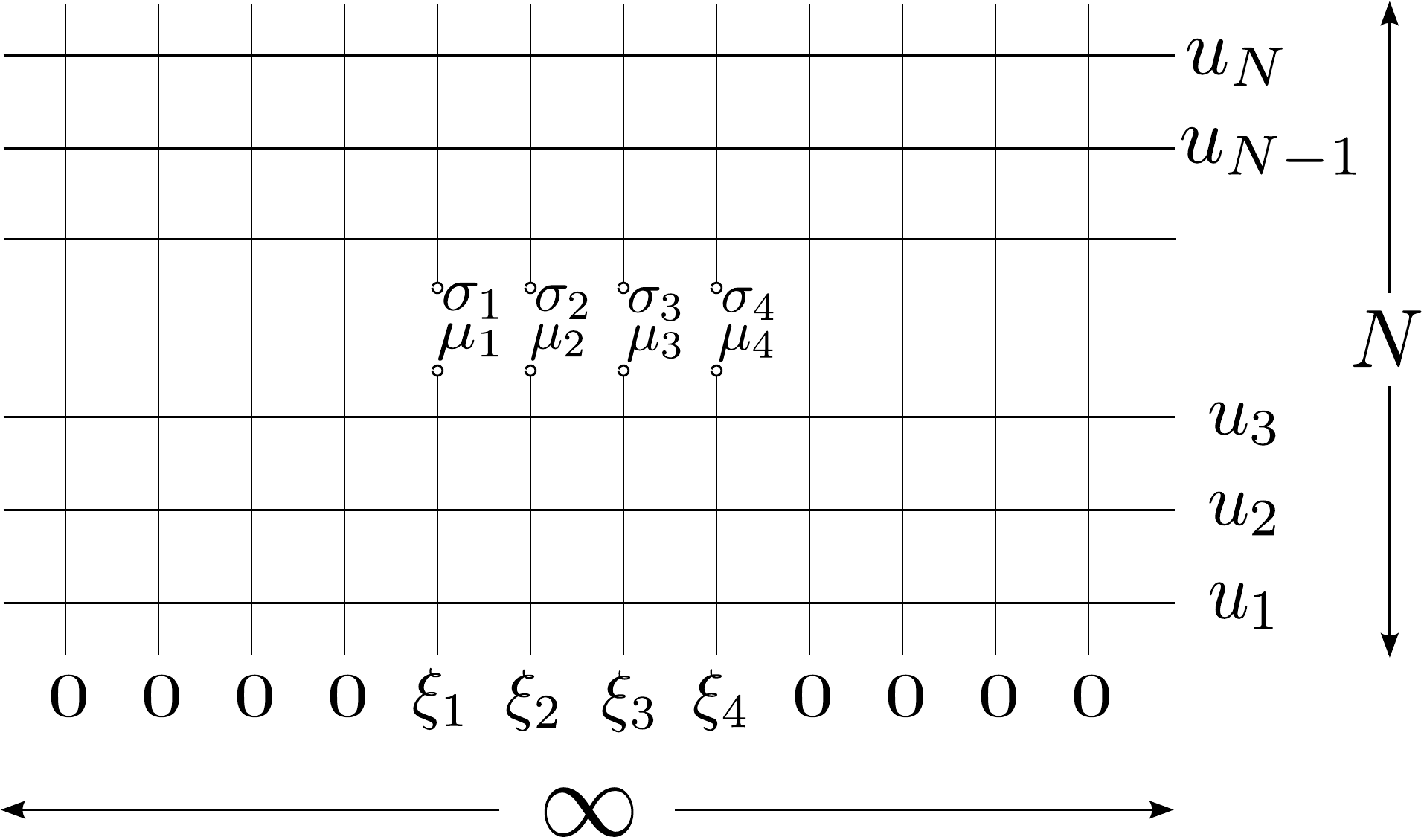}
\caption{Graphical illustration of the matrix element
  $D_{\sigma_1,...,\sigma_m}^{\mu_1,...,\mu_m}(\xi_1,...,\xi_m)$ for $m=4$.
The lattice is semi-infinite with $N$ many infinitely long horizontal lines
carrying independent spectral parameters $u_1,...,u_N$.}
\label{fig:sliced}
\end{center}
\end{figure}
The local $R$-matrix enjoys standard initial condition and crossing symmetry
which allows to perform `lattice surgery' as described in
\cite{AuKl12}. If one of the spectral
parameters $\x_1,...,\x_m$, let us say the last one $\x_m$, 
on the vertical lines with cut happens to be
identical to any of the spectral parameters on the horizontal lines, the
matrix $D(\x_1,...,\x_m)$ maps to $D(\x_1,...,\x_m-2\i)$ upon a
linear map $A$ acting on the ($2^{2m}$-dimensional) space of 
reduced density matrices.
\begin{figure}[h]
\begin{center}
\includegraphics*[width=0.95\textwidth]{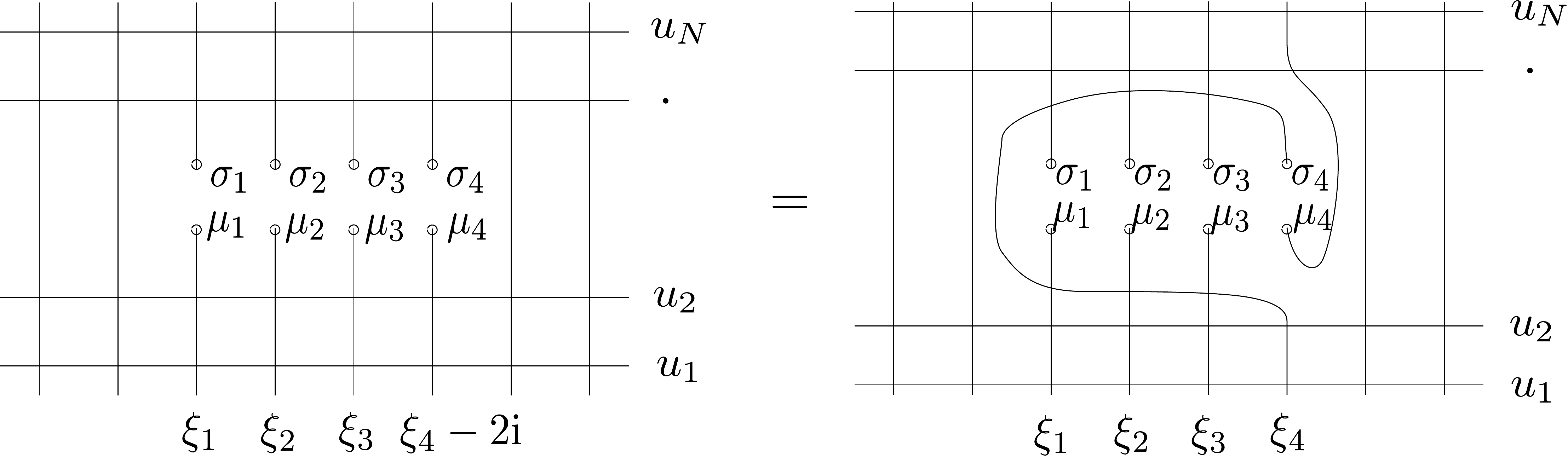}
\caption{Graphical illustration of the functional equation (\ref{Fcteq}).}
\label{fig:fcteq}
\end{center}
\end{figure}
This yields the (discrete) qKZ-type functional equation
\begin{equation}
D(\x_1,...,\x_m-2\i)=A(\x_1,...,\x_m)(\x_1,...,\x_m).\label{Fcteq}
\end{equation}
The map $A$ depends on the spectral parameters $\x_1,...,\x_m$ and is given
by a product of $2m-2$ many $R$-matrices.

In addition we have the asymptotic condition
\begin{equation}
\lim_{\x_m\to\infty}D(\x_1,...,\x_{m-1},\x_m)=D(\x_1,...,\x_{m-1})\otimes
\mathds{1}.\label{asymptotD}
\end{equation}

The under-determinacy of the functional equations for the
density matrix as function of a single variable is simply resolved by
considering the full dependence on all arguments and invoking the intertwining
(symmetry) relations
\begin{equation}
\check R_{k,k+1}(\x_k,\x_{k+1})D
(\x_1,...,\x_k,\x_{k+1},...,\x_m)\check R_{k,k+1}^{-1}(\x_k,\x_{k+1})
=
D(\x_1,...,\x_{k+1},\x_k,...,\x_m).\label{Intertw}
\end{equation}

Let us be explicit for the 2-site case which is non-trivial and will be
analyzed in detail in the next sections. 
\begin{equation}
\check R_{1,2}(\x_1,\x_2)D^{[2]}
(\x_1,\x_2)\check R_{1,2}^{-1}(\x_1,\x_2)
=D^{[2]}(\x_2,\x_1).\label{Intertw2}
\end{equation}
Here we explicitly remind of the spin-1 (level 2) case by the superscript $2$.
Due to $su(2)$ invariance some simplification occurs which is not essential
for the following reasoning, but leads to some notational simplification. The
2-site density matrix as well as the intertwiner $\check R_{1,2}$ are
superpositions of projectors onto singlet, triplet and quintuplet spaces. All
of these operators commute. Hence, $\check R_{1,2}$ and $\check R_{1,2}^{-1}$
on the \lhs\ of (\ref{Intertw2}) drop out and we obtain
$D^{[2]}(\x_1,\x_2)=D^{[2]}(\x_2,\x_1)$. Each of the entries of this density
matrix is a symmetric polynomial of two arguments of degree $2N$ (divided by
the product of eigenvalues $\La^{[2]}(\x_1)\La^{[2]}(\x_2)$). Each of these
polynomials has $(2N+1)^2$ coeffcients. Due to symmetry, only $(2N+1)(N+1)$
coefficients are independent. There are equations for $D^{[2]}(\x_1,\x_2)$ for
arbitrary first argument $\x_1$ and for $N+1$ many discrete values of the
second argument $\x_2=u_1,...,u_n, \infty$ which amount to $(2N+1)(N+1)$ many
equations for the coefficients.  (We have checked linear independence for
characteristic cases.)

The above reasoning shows that the solution to the functional equation 
(\ref{Fcteq}), asymptotics (\ref{asymptotD}) and symmetry (\ref{Intertw})
is unique for the analytic condition that each matrix element is of the type
\begin{equation}
\frac{p(\x_1,...,\x_m)}{\La^{[2]}(\x_1)...\La^{[2]}(\x_m)}
\end{equation}
where $p$ is a multivariate polynomial of degree $2N$. Finding the solution is
a different story and is based on a suitable ansatz. The result will be given
in the next sections.

The above reasoning was done for pairwise different spectral parameters on the
horizontal lines of the lattice. This allowed for a simple counting of
independent equations. In the case of degenerate values, the equations can be
formulated for the functions and for certain derivatives yielding the same
counting. In the following we are dealing with systems with $N/2$ many
spectral parameters $\i u$ and $N/2$ many
spectral parameters $2\i-\i u$. Clearly, the choice of these parameters can be
relaxed to have pairwise different numbers that still approximate the
statistical operator.

\subsection{Functional properties of $D^{[1]} (\x_1,...,\x_n)$}
\begin{figure}[h]
\begin{center}
\includegraphics*[width=0.85\textwidth]{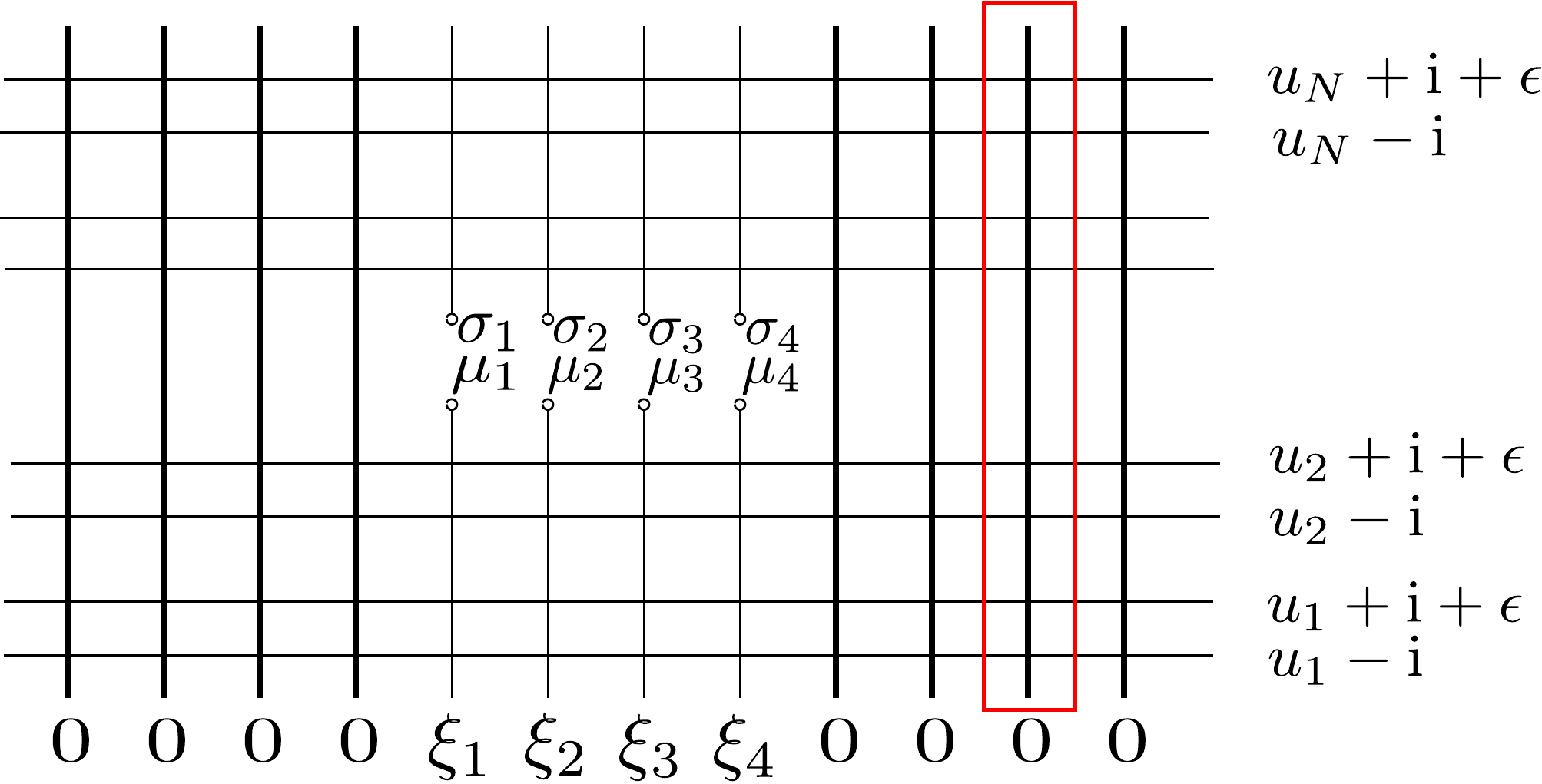}
\caption{Graphical illustration of the auxiliary lattice. If $\epsilon=0$ is
  chosen and symmetrizers at neighbouring pairs of horizontal lines with
  spectral parameters $u_j\pm\i$ are attached, the lattice will turn into a
  realization of $D^{[1]} (\x_1,...,\x_n)$. In the main text, an argument is
  given to as why the same result appears in a simple $\epsilon\to 0$ limit
  without symmetrizers.
This argument is based on the infinitely many column-to-column transfer
matrices with spin-1 auxiliary space, cf.~the boxed object in the figure.}
\label{fig:spin12}
\end{center}
\end{figure}

From a fundamental point of view it would be highly desirable to tackle the 
density matrix $D^{[1]} (\x_1,...,\x_n)$ (vertical spin-1/2 lines embedded in
spin-1 lines). There are no clear functional properties like in the $D^{[2]}$
case as here the mixed $R$-matrices do not enjoy the standard initial condition.
However, $D^{[1]} (\x_1,...,\x_n)$ may be obtained in a certain limit of a
system with (clearly) $n$ many vertical spin-1/2 lines, 
$2N$ many horizontal spin-1/2 lines and infinitely many vertical spin-1 lines,
see Fig.~\ref{fig:spin12} for the case $n=4$.
The naive idea is to use a direct fusion construction based on a lattice like
the one shown in Fig.~\ref{fig:spin12} with $\epsilon=0$ and symmetrizers
applied to the horizontal spaces. Next one would try to write down $2N$ many
functional equations of the type (\ref{Fcteq}) as now the $R$-matrices of the
vertical spin-1/2 lines and the $2N$ many horizontal spin-1/2 lines enjoy the
standard initial condition and crossing symmetry. Unfortunately, the spectral
parameters on the horizontal lines are not independent. Only $N$ many
meaningful equations can be derived, the other $N$ many attempts result in
uncontrolled $0/0$ expressions.

The system shown in Fig.~\ref{fig:spin12} is well-defined and has a regular
$\epsilon\to 0$ limit: the column-to-column transfer matrices with auxiliary
spin-1 space take their leading state close to the
space of states that are symmetric in neighbouring pairs. In the limit
$\epsilon\to 0$ the leading state converges to a state in the pure
spin-1 subspace $\left({\mathbb C}^3\right)^{\otimes N}$ of the total space
$\left({\mathbb C}^2\right)^{\otimes 2N}$. This is clear from the root pattern
of the Bethe ansatz solution to the system Fig.~\ref{fig:spin12}.

The benefit of dealing with the `$\epsilon$-regularized' system is the
possibility to now formulate $2N$ many functional equations. These equations
are valid for any small, but finite $\epsilon$ for the semi-infinite
system. This is the program of the pure spin-1/2 case that leads to algebraic
expressions of the density matrix in terms of nearest-neighbour correlators
\cite{JMS08,AuKl12}. The algebraic structure is completely
independent of $\epsilon$! Just the nearest-neighbour correlators depend on
$\epsilon$, but have a regular limit. We therefore see that the factorized
expressions also hold for the object $D^{[1]} (\x_1,...,\x_n)$ which is of
interest to us. Now the computational strategy is to write down the entries of 
$D^{[1]} (\xx_1,...,\xx_n)$ for $n=2m$ in terms of the fundamental nearest
neighbour function 
$\Omega(\x_1,\x_2)$ that is calculated in the next section. Then the
symmetrization in the quantum spaces (vertical lines) is to
be performed for $\xx_{2k-1}=\x_k-\i,\xx_{2k}=\x_k+\i$ ($k=1,...,m$).

\subsection{Bethe Ansatz solution}
For the calculation of the free energy (\ref{free}) and the inhomogeneous
density matrix (\ref{d2inhom}) we need to know in first place the
dominant eigenvector $|\Ps\>$ and the corresponding transfer matrix
eigenvalue $\La^{[2]} (\la)$. They can be obtained by means of the
standard algebraic Bethe ansatz for the spin-$\2$ generalized model
(see e.g.\ chapter 12.1.6 of \cite{thebook}), since, by the general
reasoning of the fusion procedure \cite{KuSk82b}, the quantum transfer
matrix $t^{[2]} (\la)$ can be expressed in terms of a transfer matrix
with spin-$\2$ auxiliary space.

For our purposes we consider a staggered monodromy
matrix with spin-$\2$ auxiliary space and $N$ many spin-1 quantum spaces and
two spin-1/2 spaces by
\begin{multline} \label{defmono12}
     T_a^{[1]} (\la + \i) = \re^{\beta h \s_a^z}
        R_{a, N+2}^{[1,1]} (\la -\mu+\i)
        R_{N+1, a}^{[1,1] \: t_1} (\mu+\delta-\la-\i)\cdot\\
\cdot        R_{a, N}^{[1,2]} (\la +\i u)
        R_{N-1, a}^{[2,1] \: t_1} (\i u - \la) \dots
        R_{a, 2}^{[1,2]} (\la +\i u)
        R_{1, a}^{[2,1] \: t_1} (\i u - \la) \epp
\end{multline}
Then, interpreting this monodromy matrix as a $2 \times 2$ matrix in
the auxiliary space $a$, we define the transfer matrix and the quantum determinant
(spin-0 fusion) as
\begin{equation} \label{deft1detq1}
     t^{[1]} (\la) = \tr T^{[1]} (\la) \epc \qd
     \detq T^{[1]} (\la) = U \bigl( T^{[1]} (\la - \i)
                               \otimes T^{[1]} (\la + \i) \bigr) U^t \epc
\end{equation}
where $U$ is the projector onto the antisymmetric state in the tensor product of
the two auxiliary spaces.
It follows from (\ref{fusedr}) that
\begin{equation} \label{spin1mono}
     T^{[2]} (\la) = S \bigl( T^{[1]} (\la - \i)
                              \otimes T^{[1]} (\la + \i) \bigr) S^t \epc
\end{equation}
where $S$ is the projector onto symmetric states in the tensor product of the
two auxiliary spaces.
Taking the trace and using (\ref{deft1detq1}) we conclude that
\begin{equation} \label{t2t1qdet}
     t^{[2]} (\la) = t^{[1]} (\la - \i) t^{[1]} (\la + \i)
                     - \detq T^{[1]} (\la) \epc
\end{equation}
Hence, since $\detq T^{[1]} (\la)$ commutes with $T^{[1]} (\la)$ 
\cite{KuSk82b}, every eigenstate of $t^{[1]} (\la)$ is an eigenstate of
$t^{[2]} (\la)$ as well.

The algebraic Bethe ansatz is based on the Yang-Baxter algebra relations
\begin{equation} \label{yba}
     \check R^{[1,1]} (\la - \m)
        \bigl( T^{[1]} (\la) \otimes T^{[1]} (\m) \bigr) =
        \bigl( T^{[1]} (\m) \otimes T^{[1]} (\la) \bigr)
        \check R^{[1,1]} (\la - \m)
\end{equation}
which follow from (\ref{ybes}) and (\ref{defmono12}). Representing
$T^{[1]} (\la)$ by the $2 \times 2$ matrix
\begin{equation}
     T^{[1]} (\la) = \begin{pmatrix}
                        A(\la) & B(\la) \\ C(\la) & D(\la)
                     \end{pmatrix},
\end{equation}
the quantum determinant is
\begin{equation} \label{simpleqdet}
     \detq T^{[1]} (\la) = D(\la-\i) A(\la+\i) - B(\la-\i) C(\la+\i) \epp
\end{equation}
Defining the pseudo vacuum
\begin{equation}
     |0\> =
                \Bigl[
        \Bigl( \begin{smallmatrix} 0 \\ 0 \\ 1 \end{smallmatrix} \Bigr)
        \otimes
        \Bigl( \begin{smallmatrix} 1 \\ 0 \\ 0 \end{smallmatrix} \Bigr)
        \Bigr]^{\otimes \frac{N}{2}}
\otimes
\bigl( \begin{smallmatrix} 0 \\ 1 \end{smallmatrix} \bigr) \otimes
\bigl( \begin{smallmatrix} 1 \\ 0 \end{smallmatrix} \bigr)
\end{equation}
we deduce from (\ref{defmono12}) that
\begin{equation}
     C(\la) |0\> = 0 \epc \qd
     A(\la) |0\> = a(\la) |0\> \epc \qd D(\la) |0\> = d(\la) |0\> \epc
\end{equation}
where the pseudo vacuum eigenvalues $a(\la)$ and $d(\la)$ are explicit
complex valued functions. Using the notation
\begin{equation} \label{defphipm}
     \phi_\pm (\la) = (\la \pm \i u)^{N/2} \epc \qd u = - \frac{J}{NT},
\end{equation}
which proved to be useful in \cite{Suzuki99}, and
\begin{equation} \label{defphipm2}
\vphi_+(\la):=\la-\mu,\qquad \vphi_-(\la):=\la-\mu-\delta,\ 
\end{equation}
which arise from the additional horizontal spin-1/2 lines carrying the spectral 
parameters $\mu-\i$ and $\mu-\i+\delta$ (at the vertex with transposition),
we can express the vacuum eigenvalues as
\begin{equation} \label{vacexp}
a(\la)=\re^{h/T}\frac{\phi_-(\la+\i)}{\phi_-(\la-3\i)}
\frac{\vphi_-(\la)}{\vphi_-(\la-2\i)}\epc\qd
d(\la)=\re^{-h/T}\frac{\phi_+(\la-\i)}{\phi_+(\la + 3 \i)}
\frac{\vphi_+(\la)}{\vphi_+(\la + 2 \i)}\epp
\end{equation}

Given the Yang-Baxter algebra (\ref{yba}) and the pseudo vacuum
eigenvalues (\ref{vacexp}) the eigenvectors and eigenvalues of
$t^{[1]} (\la)$ can be obtained from general considerations (see e.g.\
chapter 12.1.6 of \cite{thebook}). The dominant eigenstate $|\Ps\>$
of $t^{[2]} (\la)$, in particular, can be represented as
\begin{equation} \label{domi}
     |\Ps\> = B(\la_1) \dots B(\la_{N+1}) |0\> \epc
\end{equation}
where the set of so-called Bethe roots $\{\la_j\}_{j=1}^{N+1}$ is a specific
solution of the Bethe ansatz equations
\begin{equation} \label{bae}
      \frac{a(\la_j)}{d(\la_j)} = \prod_{\substack{k = 1 \\ k \ne j}}^{N+1}
         \frac{\la_j - \la_k + 2\i}{\la_j - \la_k - 2\i} \epc \qd
         j = 1, \dots, N+1 \epp
\end{equation}
For the given set of Bethe roots $\{\la_j\}_{j=1}^{N+1}$ we define the
$Q$-function
\begin{equation} \label{defq}
     q(\la) = \prod_{j=1}^{N+1} (\la - \la_j) \epp
\end{equation}
Then the eigenvalue of $t^{[1]} (\la)$ corresponding to $|\Ps\>$ is
\begin{equation}
     \La^{[1]} (\la) = a(\la) \frac{q(\la - 2 \i)}{q(\la)}
          + d(\la) \frac{q(\la + 2 \i)}{q(\la)} \epp
\end{equation}
As for the eigenvalue of $t^{[2]} (\la)$ we conclude with (\ref{t2t1qdet})
and equation (\ref{simpleqdet}) above that
\begin{equation} \label{eva}
     \La^{[2]} (\la) = \La^{[1]} (\la - \i) \La^{[1]} (\la + \i)
                         - a(\la + \i) d(\la - \i)
\end{equation}
This eigenvalue and the Bethe ansatz equations (\ref{bae}) are the main
input for the calculation of the thermodynamics of the spin-1
chain. In order to perform the Trotter limit the eigenvalue must
be represented by means of auxiliary functions satisfying a finite
set of nonlinear integral equations. This was achieved in
\cite{Suzuki99}.
\section{Integral equations and basic functions}
\label{sec:therm}
In this section we consider the evaluation of the largest eigenvalue of the
generalized quantum transfer matrix by means of nonlinear integral equations
(NLIE). This gives us the opportunity to introduce certain auxiliary functions
that are also relevant for the factorized algebraic expressions of the density
matrix elements in the next section. Our starting point is the expression for
the dominant eigenvalue $\La^{[2]} (\la)$ of the quantum transfer matrix
together with the Bethe ansatz solution (\ref{bae})-(\ref{eva}). In
\cite{Suzuki99} the problem was solved within the more general context of the
fusion hierarchy, and NLIE for the integrable isotropic spin chains of
arbitrary spin were obtained. 

For the calculation of the free energy for spin 1 we will be dealing with
three coupled NLIE for three functions $\fb$, $\fbq$ and $y$. We show the
equations below in (\ref{nlie}) and present the derivation in
appendix~\ref{app:contnlietonlie}. (Most of the relations of this section are
valid for arbitrary magnetic field $h$, however the main applications later on
concern the zero field case.) For finite Trotter number $N$ the
functions $\fb$, $\fbq$ and $y$ can be expressed in terms of the
$Q$-function (\ref{defq}) and the functions $\Ph_\pm$ introduced in
(\ref{defphipm}) (see appendix~\ref{app:auxfun}). This defines them as
meromorphic functions on the entire complex plane, but is inappropriate
for performing the Trotter limit. In the NLIE, on the other hand, the
Trotter number appears only in the driving term and the Trotter limit
is easily obtained. For a discussion of some of the subtleties related
to the Trotter limit and the definition of useful auxiliary functions
see \cite{GoSu10}.

The functions $\fb(\la), \fbq(\la)$
and $y(\la)$ are defined in (\ref{auxbbq}), (\ref{yY}) and will
be of particular use in the neighbourhood of the real axis.
For convenience we introduce the shifted functions
\begin{equation}
    \fb_{\epsilon}(\la)=\fb(\la-\i\epsilon) \epc \quad
    \fbq_{\epsilon}(\la)=\fbq(\la+\i\epsilon) \epc
\end{equation}
and similar capital functions, $\Bf_{\epsilon}(\la) = 1 + \fb_{\epsilon}(\la)$ 
etc. Then the desired NLIE read
\begin{equation} \label{nlie}
     \begin{pmatrix}
        \log y(\la) \\
	\log \fb_{\epsilon} (\la) \\
	\log \fbq_{\epsilon}(\la)
     \end{pmatrix} = 
     \begin{pmatrix}
        \Delta_y(\la)\\
	\Delta_b(\la) \\
	\Delta_{\overline{b}}(\la)
     \end{pmatrix} +
     \widehat{\cal K} * 
     \begin{pmatrix}
        \log Y(\la) \\
	\log \Bf_{\epsilon} (\la) \\
	\log \Bfq_{\epsilon} (\la)
     \end{pmatrix} \epc
\end{equation}
where $(\widehat{\cal K}*g)_i$ denotes the matrix convolution
$\sum_j \int_{-\infty}^{\infty} \rd \la'\, \widehat{\cal K}_{i,j}(\la-\la')
g_j (\la')$, and
\begin{subequations}
\begin{align}
     \Delta_y(\la)&=\log\left[\frac{\tanh\left(\frac\pi 4(\la-\mu+\i)\right)}
{\tanh\left(\frac\pi 4(\la-\mu-\delta+\i)\right)}\right]\simeq -\i
\frac{\frac\p 2}{\cosh \frac\p 2 (\la-\mu)} \cdot\delta,\\
     \Delta_b(\la)
         &= - \frac{h}{T} + d(u,\la - \i \eps) \epc \qd
	  \Delta_{\overline{b}}(\la) = \frac{h}{T} + d(u,\la + \i \eps)
	  \epc \\[1ex]
     d(u,\la) & = \frac{N}{2}
        \int_{-\infty}^{\infty} \rd k\, \re^{- \i k \la}
	\frac{\sinh uk}{k\cosh k}\
	\overset{N \rightarrow \infty}{\longrightarrow}\
	- \frac{J}{T} \frac{\frac\p 2}{\cosh \frac\p 2 \la} \epp
\end{align}
\end{subequations}
The integration constants ($\pm h/T$) are fixed by comparing the
asymptotic values of both sides of (\ref{nlie}) for $|\la| \rightarrow
\infty$. 
The kernel matrix is given by
\begin{equation} \label{ksym}
     \widehat{\cal {\cal K}}(\la) =
        \begin{pmatrix}
	   0 & {\cal K}(\la+\i\epsilon) & {\cal K}(\la-\i\epsilon) \\
	   {\cal K}(\la-\i\epsilon) & {\cal F}(\la) &
           -{\cal F}(\la+2\i(1-\epsilon)) \\
           {\cal K}(\la+\i\epsilon) & -{\cal F}(\la-2\i(1-\epsilon)) &
           {\cal F}(\la)
        \end{pmatrix} \epc
\end{equation}
where  
\begin{equation}
     {\cal K}(\la) = \frac{1}{4 \cosh\pi \la/2} \epc \qd
     {\cal F}(\la) = \int_{-\infty}^{\infty} \frac{\rd k}{2 \p}\,
                 \frac{e^{-|k|- \i k\la}}{2 \cosh k} \epc\label{kernelKF}
\end{equation}
and ${\cal F}$ can be expressed in terms of the digamma function $\Psi$, see
(\ref{Fexplicit}).

The derivation of integral expressions for the eigenvalues is
involved, we defer the details to appendix~\ref{app:contnlietonlie}.
The results are
\begin{subequations}
\begin{multline}
\log\Lambda^{[1]}(\la)= 
-\int_{-\infty}^{\infty}{dk\:\frac{N
    e^{-3\left|k\right|}\cosh\left((u+1)k\right)}{2\left|k\right|\cosh\left(k\right)}e^{-\i
    k\la}}-\log\left[\phi_{+}\left(\la+3\i\right) \phi_{-}\left(\la-3\i\right)\right]+\text{cst.}\\
+\int_{-\infty}^{\infty}{d\la'\:\mathcal{K}(\la-\la')}\biggl\{\log Y(\la')
+\log\left[\frac{\vphi_+\left(\la-\i\right)\vphi_-\left(\la+\i\right) }{
    \vphi_{-}\left(\la-\i\right) \vphi_{+}\left(\la+\i\right) }\right]
\biggr\}
\end{multline}
respectively
\begin{align}
\log\Lambda^{\left[2\right]}(\la)&=
\log\left[\frac{\phi_+\left(\la-2\i\right)\phi_-\left(\la+2\i\right) }
  {\phi_{-}\left(\la-2\i\right) \phi_{+}\left(\la+2\i\right)}
\frac{\vphi_+\left(\la-\i\right)\vphi_-\left(\la+\i\right) }{
    \vphi_{-}\left(\la-\i\right) \vphi_{+}\left(\la+\i\right) }
\right]
+\log y(\la)
\end{align}
\end{subequations}

Next we are interested in
$\frac{\partial}{\partial\delta}\log\Lambda^{[1]}$ for
which we derive an integral expression in terms of functions satisfying a set
of linear integral equations. The basic functions are
the derivatives of the functions $\log Y, \log \Bf_{\epsilon}, \log
\Bfq_{\epsilon}$ with respect to $\delta$. We have the identities
\begin{subequations}\label{Gfunction}
\begin{align}
G_{y}(\la,\mu)&:=\left.\frac{\partial}{\partial\delta}\log
  Y\left(\la\right)\right|_{\delta=0}\ \text{,} 
& \left.\frac{\partial}{\partial\delta}\log y\left(\la\right)\right|_{\delta=0}=[1+y^{-1}(\la)]|_{\delta=0}G_{y}(\la,\mu)\ \text{,}\\
G_{\fb}(\la,\mu)&:=\left.\frac{\partial}{\partial\delta}\log
\Bf_{\epsilon}\left(\la\right)\right|_{\delta=0}\ \text{,} 
& \left.\frac{\partial}{\partial\delta}\log \fb_{\epsilon}\left(\la\right)\right|_{\delta=0}=[1+\fb_{\epsilon}^{-1}(\la)]|_{\delta=0}G_{\fb}(\la,\mu)\ \text{,}\\
G_{\fbq}(\la,\mu)&:=\left.\frac{\partial}{\partial\delta}\log
\Bfq_{\epsilon}\left(\la\right)\right|_{\delta=0}\ \text{,} 
& \left.\frac{\partial}{\partial\delta}\log \fbq_{\epsilon}\left(\la\right)\right|_{\delta=0}=[1+\fbq_{\epsilon}^{-1}(\la)]|_{\delta=0}G_{\fbq}(\la,\mu)\ \text{.}
\end{align}
\end{subequations}
These satisfy the set of linear integral equations
\begin{equation}\label{ErgebnisGFkten}
\begin{pmatrix}
[1+y^{-1}(\la)]G_{y}(\la,\mu) \\ [1+\mathfrak{b}^{-1}_{\epsilon}(\la)]G_{\fb}(\la,\mu) \\ [1+\overline{\mathfrak{b}}^{-1}_{\epsilon}(\la)]G_{\fbq}(\la,\mu)
\end{pmatrix} = 
\begin{pmatrix}
-\i\frac{\frac\pi 2}{\cosh\frac{\pi}{2}(\la-\mu)} \\ 0 \\ 0
\end{pmatrix} + \widehat{\mathcal{K}}*
\begin{pmatrix}
G_{y}(\la,\mu) \\ G_{\fb}(\la,\mu) \\ G_{\fbq}(\la,\mu)
\end{pmatrix}\ \text{,}
\end{equation}
Next we obtain an explicit expression of the eigenvalue's derivative in terms
of the $G$ functions
\begin{subequations}
\begin{align}
\left.\frac{\partial}{\partial\delta}\log\Lambda^{[1]}(\la)\right|_{\delta=0}&=
\int_{-\infty}^{\infty}{d\la'\:\frac{1}{4\cosh\left(\frac{\pi}{2}(\la-\la')\right)}}\left\{
G_{y}(\la',\mu)+\frac{2\i}{(\la'-\mu)^2+1}\right\}\label{NLIESpin1.2a}\\
&=\int_{-\infty}^{\infty}{d\la'\:\frac{1}{4\cosh\left(\frac{\pi}{2}(\la-\la')\right)}} G_{y}(\la',\mu)+2\pi i {\cal F}(\la-\mu) \text{,}\label{NLIESpin1.2b}
\end{align} 
\end{subequations}
for $|\operatorname{Im}(\la)|<1$.
Next we note a useful identity for ${\cal F}$ defined in (\ref{kernelKF})
\begin{multline}
{\cal F}(\la)
=\frac{1}{8\pi}\Biggl\{\Psi\left(-\frac i 4\la\right)+\Psi\left(\frac i 4\la\right)-\Psi\left(\frac12-\frac i 4\la\right)-\Psi\left(\frac12+\frac i 4\la\right)\Biggr\},\label{Fexplicit}
\end{multline}
where $\Psi$ is the standard digamma function 
\begin{equation}
\Psi(x):=\frac{\partial}{\partial x}\log(\Gamma(x))\ \text{, }\quad 
\Gamma(x):= \int\limits_0^\infty t^{x-1} { e}^{-t} dt \ \text{.}
\end{equation}
We remind of the functional equations
\begin{equation}
\Psi(1-x)=\Psi(x)+\pi\cot(\pi x)\ \text{, }\ x\neq\frac12\ \text{,}
\end{equation}
and
\begin{equation}
\Psi(x+1)=\Psi(x)+\frac{1}{x}\ \text{.}
\end{equation}
The eigenvalue's derivative satisfies the following 2-point equation
\begin{equation}
\left.\frac{\partial}{\partial\delta}\ln\Lambda^{[1]}(\la+\i)\right|_{\delta=0}+\left.\frac{\partial}{\partial\delta}\ln\Lambda^{[1]}(\la-\i)\right|_{\delta=0}= G_{y}(\la,\mu) +\frac{1}{\la-\mu-\i}-\frac{1}{\la-\mu+\i}\ \text{.}
\end{equation}
The \rhs\ does not simplify in any obvious way to elementary expressions due
to the occurrence of $G_{y}(\la,\mu)$. In the next subsection we will derive a 
simple 3-point equation involving just derivatives of $\Lambda^{[1]}$.

Due to $su(2)$ invariance the density operator $D^{[1]}(\la,\mu)$ for
two neighbouring spin-1/2 spaces with spectral parameters $\la, \mu$ 
has the following representation
\begin{equation}\label{omegaDarstellungDichte1}
D^{[1]}(\la,\mu)=\left(\frac14-\frac16\omega(\la,\mu)\right)\mathds{1}+\frac13 \omega(\la,\mu) P^{[1]}
\end{equation}
where $P^{[1]}$ is the permutation operator of neighbouring spin-1/2 objects
and $\omega(\la,\mu)$ is some (symmetric)
function. In Appendix \ref{app:simplcorr} we show that
\begin{equation}
 \omega(\la,\mu) = \frac12-\frac{(\la-\mu)^2+4}{2\i}\:\frac{\partial}{\partial\delta}\left.\ln\left\{\Lambda^{[1]}(\la;\mu)\right\}\right|_{\delta=0}\ \text{.}
 \label{Integraldarstellungomega}
\end{equation}
For later purposes, it is convenient to use another closely related function
\begin{align}
\Omega(\la,\mu)&:=\frac{2\i}{(\la-\mu)^2+4)}\operatorname{tr}\left\{D^{[1]}(\la,\mu)\:P^{[1]}\right\}\nonumber\\
&=2\i\frac{\omega(\la,\mu)+\frac 12}{(\la-\mu)^2+4}=-\frac{\partial}{\partial\delta}\left.\ln\left\{\Lambda^{[1]}(\la;\mu)\right\}\right|_{\delta=0}+\frac{2\i}{(\la-\mu)^2+4}
\label{omegaTransfermatrix}
\end{align}
An interesting and useful observation concerns the trace of $P^{[1]}$ over
the symmetric subspace for arguments $(\la,\mu)$ chosen as $(\la-\i,\la+\i)$
which according to the fusion principles yields 
\begin{equation}
\frac 34+\frac 12\omega(\la-\i,\la+\i)=\frac{\La_2(\la)}{\La_1 (\la+\i)\La_1
  (\la-\i)}=\frac 1{1+y(\la)^{-1}}.\label{Lambdaratio}
\end{equation}
We have the explicit expression (\ref{Integraldarstellungomega}) which seems
to yield $\frac 12$ for $\omega(\la-\i,\la+\i)$. However,
$\frac{\partial}{\partial\delta}\ln\Lambda^{[1]}(\la;\mu)$
develops poles for $\la-\mu=\pm 2\i$ which can be evaluated on the basis of 
(\ref{NLIESpin1.2a}) and (\ref{ErgebnisGFkten}) in full agreement with
(\ref{Lambdaratio}).

\subsection{Functional equations for the basic functions}
From the definition (\ref{lambdapol1}), (\ref{lambdapol2}) of the functions $\La_1$ and $\La_2$ 
we find the relations
\begin{align}
\La_1 (\la-\i)\La_1 (\la+\i)&=\phi(\la - 3 \i) \phi(\la + 3 \i) 
\vphi(\la - 2\i) \vphi(\la + 2 \i)+ \La_2 (\la)\nonumber\\ 
&=\phi(\la - 3 \i) \phi(\la + 3 \i)
\vphi(\la - 2 \i) \vphi(\la + 2 \i)\cdot Y(\la)\label{Lafctnal1}\\ 
\La_2 (\la-\i)\La_2(\la+\i)&=
\phi(\la - 2 \i)\phi(\la - 4 \i)\phi(\la +4 \i) \phi(\la + 2 \i)\times\nonumber\\
&\times\vphi(\la - \i) \vphi(\la -3 \i)\vphi(\la +3 \i) \vphi(\la + \i)+
\phi(\la)\tilde\La (\la).\label{Lafctnal2}
\end{align}
We divide the first equation by $\vphi(\la - 2\i) \vphi(\la + 2 \i)$ and
the second equation by a product of that function with $\la$ replaced by
$\la\pm\i$
\begin{align}
\frac{\La_1 (\la-\i)\La_1 (\la+\i)}{\vphi(\la - 2\i) \vphi(\la + 2 \i)}
&=\phi(\la - 3 \i) \phi(\la + 3 \i) 
+ \frac{\La_2 (\la)}{\vphi(\la - 2\i) \vphi(\la + 2 \i)}\label{Lafctnal1a}\\ 
\frac{\La_2 (\la-\i)}{\vphi(\la -3 \i)\vphi(\la +\i)}\cdot
\frac{\La_2(\la+\i)}{\vphi(\la -\i)\vphi(\la +3\i)}&=
\phi(\la - 2 \i)\phi(\la - 4 \i)\phi(\la +4 \i) \phi(\la + 2 \i)
+\nonumber\\
&+\frac{\phi(\la)}{\vphi(\la - \i) \vphi(\la -3 \i)\vphi(\la +3 \i) \vphi(\la + \i)}\tilde\La (\la).\label{Lafctnal2a}
\end{align}
From the last equation we see that
\begin{equation}\label{Lafctnal2b}
\frac{\partial}{\partial\delta}\ln\frac{\La_2 (\la-\i)}{\vphi(\la -3
  \i)\vphi(\la +\i)}
+
\frac{\partial}{\partial\delta}\ln\frac{\La_2(\la+\i)}{\vphi(\la -\i)\vphi(\la
  +3\i)}
=0\quad\hbox{if}\ \phi(\la)=0,
\end{equation}
since for $\la$'s with $\phi(\la)=0$ the \rhs\ of (\ref{Lafctnal2a}) is
absolutely independent of $\delta$.
\begin{align}\label{Lafctnal2c}
\frac{\partial}{\partial\delta}\ln\frac{\La_2 (\la)}{\vphi(\la - 2\i)
  \vphi(\la + 2 \i)}
&=\frac{\vphi(\la - 2\i) \vphi(\la + 2 \i)}{\La_2 (\la)}
\frac{\partial}{\partial\delta}\frac{\La_2 (\la)}{\vphi(\la - 2\i) \vphi(\la +
  2 \i)}\nonumber\\
&=\frac{\vphi(\la - 2\i) \vphi(\la + 2 \i)}{\La_2 (\la)}
\frac{\partial}{\partial\delta}
\frac{\La_1 (\la-\i)\La_1 (\la+\i)}{\vphi(\la - 2\i) \vphi(\la + 2
  \i)}\nonumber\\
&=\frac{\La_1 (\la-\i)\La_1 (\la+\i)}{\La_2 (\la)}
\frac{\partial}{\partial\delta}\ln
\frac{\La_1 (\la-\i)\La_1 (\la+\i)}{\vphi(\la - 2\i) \vphi(\la + 2 \i)},
\end{align}
where for the second identity we have used (\ref{Lafctnal1a}) where the first
summand on the \rhs\ is independent of $\delta$. Next, inside the logarithm, 
we replace $\La_1$ by
$\La^{[1]}$ and the logarithmic derivative by the function $\Omega$ 
(\ref{omegaTransfermatrix}). We then find for the \rhs\ of (\ref{Lafctnal2c})
\begin{equation}\label{Lafctnal2d}
...=\frac{\La_1 (\la-\i)\La_1 (\la+\i)}{\La_2 (\la)}
\left(\frac{3\i}{(\la-\mu)^2+9}-\frac{\i}{(\la-\mu)^2+1}
-\Omega(\la-\i,\mu)-\Omega(\la+\i,\mu)\right).
\end{equation}
The ratio of $\La$ functions can be simplified by use of (\ref{Lambdaratio}). With
\begin{equation}
N(\xi):=\frac 34+\frac 12\omega(\xi-\i,\xi+\i),
\quad
o(\la):=  \frac{2\i(\la^2-3)}{(\la-3\i)(\la-\i)(\la+\i)(\la+3\i)}.
\label{defNo}
\end{equation}
we obtain
\begin{equation}\label{Lafctnal2e}
\frac{\partial}{\partial\delta}\ln\frac{\La_2 (\la)}{\vphi(\la - 2\i)
  \vphi(\la + 2 \i)}
=
-\frac1{N(\la)}\left(\Omega(\la-\i,\mu)+\Omega(\la+\i,\mu)
-o(\la-\mu)\right).
\end{equation}
Finally, the functional equation (\ref{Lafctnal2b}) takes the form
\begin{equation}\label{Dreipunkt}
\frac{\Omega(\la-2\i,\mu)+\Omega(\la,\mu)
-o(\la-\mu-\i)}{N(\la-\i)}+
\frac{\Omega(\la,\mu)+\Omega(\la+2\i,\mu)
-o(\la-\mu+\i)}{N(\la+\i)}=0,
\end{equation}
for all $\la$ being zeros of $\phi(\la)$.

\subsection{Low and high temperature limits}
{\em Low temperature limit, zero field}

From \eqref{nlie} we read off for zero field $h=0$
the zero-temperature limit $T\to 0$ 
of the auxiliary functions $y$, $\fb_{\epsilon}$ and
$\fbq_{\epsilon}$
\begin{equation}
\lim_{T\to0} \left.y\right|_{\delta=0} =1\ ,\ \lim_{T\to0} \left.\fb_{\epsilon}\right|_{\delta=0}=0=\lim_{T\to0} \left.\fbq_{\epsilon}\right|_{\delta=0}\ .
\end{equation}
And from (\ref{ErgebnisGFkten}) we obtain in the same limit
\begin{equation}
G_{y}(\la,\mu)=-\frac{\i}2\frac{\frac\pi 2}{\cosh\frac{\pi}{2}(\la-\mu)},\quad
G_{\fb}(\la,\mu) =G_{\fbq}(\la,\mu)=0.
\end{equation}
Hence the zero-temperature limit of the functions $\Omega$ and $\omega$
\eqref{omegaTransfermatrix} is calculated directly from
\eqref{NLIESpin1.2b} yielding
\begin{multline}\label{omegazerolimit}
\lim_{T\to0}\omega(\la,\mu) 
=\frac12-\frac{1}{8}((\la-\mu)^2+4)\Biggl(  \Biggl\{\Psi\left(-\frac i 4(\la-\mu)\right)+\Psi\left(\frac i 4(\la-\mu)\right)\\-\Psi\left(\frac12-\frac i 4(\la-\mu)\right)-\Psi\left(\frac12+\frac i 4(\la-\mu)\right) \Biggr\} -\frac{\pi(\la-\mu)}{2\sinh\left(\frac{\pi}{2}(\la-\mu)\right)} \Biggr)\ .
\end{multline}

\noindent
{\em High temperature expansion, zero field}

We expand the auxiliary functions in powers of $1/T$
\begin{subequations}\label{HochtempNLIE}
\begin{align}
\left.y(\la)\right|_{\delta=0}=y_0+\sum_{k=1}^{\infty}\left(\frac{1}{T}\right)^k y_k(\la)\ ,\\
\left.\fb_{\epsilon}(\la)\right|_{\delta=0}=\fb_0+\sum_{k=1}^{\infty}\left(\frac{1}{T}\right)^k \fb_k(\la)\ ,\\
\left.\fbq_{\epsilon}(\la)\right|_{\delta=0}=\fbq_0+\sum_{k=1}^{\infty}\left(\frac{1}{T}\right)^k \fbq_k(\la)\ ,
\end{align}
\end{subequations}
and solve the NLIE iteratively. The $0^{\rm th}$ order is solved by the
constants $y_0=3$, $\fb_0=\fbq_0=2$. The NLIE for the auxiliary functions up
to $1^{\rm st}$ order linearize in $y_1(\la)$, $\fb_1(\la)$, $\fbq_1(\la)$ and
can be solved by Fourier transformation. This can be repeated over and over
again: the NLIE linearize in the next (unknown) order where the (known) lower
order terms appear in non-linear combinations as driving terms.  We restrict
ourselves to the first order calculation
\begin{subequations}
\begin{align}
\left.y(\la)\right|_{\delta=0}&=3-\frac{128}{(4+\la^2)(16+\la^2)}\frac JT+\mathcal{O}(1/T^2)\ ,\\
\left.\fb_{\epsilon}(\la)\right|_{\delta=0}&=2-\frac{32\i(4+\i(\la-\i\epsilon)+(\la-\i\epsilon)^2)}{((\la-\i\epsilon)^2+1)((\la-\i\epsilon)^2+9)(\la-\i\epsilon+5\i)}\frac
JT+\mathcal{O}(1/T^2)\ ,\\
\left.\fbq_{\epsilon}(\la)\right|_{\delta=0}&=2+\frac{32\i(4-\i(\la+\i\epsilon)+(\la+\i\epsilon)^2)}{((\la+\i\epsilon)^2+1)((\la+\i\epsilon)^2+9)(\la+\i\epsilon-5\i)}\frac
JT+\mathcal{O}(1/T^2)\ .
\end{align}
\end{subequations}
Using this strategy also for the auxiliary functions $G_{y}$, $G_{\fb}$,
$G_{\fbq}$, the high
temperature expansion for the logarithmic derivative of the eigenvalue with
respect to $\delta$ leads to
\begin{equation}
\left.\frac{\partial}{\partial \delta}\log\La^{\left[1\right]}(\la)\right|_{\delta=0} = \frac{\i}{(\la-\mu)^2+4}\Biggl\{1+\frac{32}{(9+\la^2)(9+\mu^2)}\frac{J}{T} \Biggr\}+\mathcal{O}\left(1/T^2\right).
\end{equation}
Thus we have
\begin{equation}\label{omegahighlimit}
\omega(\la,\mu)=-\frac{16}{(9+\la^2)(9+\mu^2)}\frac{J}{T}+\mathcal{O}(1/T^2)\ .
\end{equation}

\section{Computation of the 2-site density matrix}
\label{sec_2sites}
Due to rotational invariance the density operator of two neighbouring
spin-1 sites has to take the form
\begin{equation}\label{2PlatzDichteMatrix}
D^{[2]}(\xi_1,\xi_2)=\rho_{1}(\xi_1,\xi_2)\:\mathds{1}+\rho_{2}(\xi_1,\xi_2)\:P^{[2]}+\rho_{3}(\xi_1,\xi_2)\:P^{(0)},
\end{equation}
where $P^{[2]}$ is the permutation operator and $P^{(0)}:=(3\times$) the 
projector onto the singlet. Three coefficients $\rho_{k}$, $k=1,2,3$, 
have to be determined.

The general functional equation of qKZ type specialized to the 2-site case and
written in terms of the above introduced coefficients reads
\begin{equation}\label{Fktglabc}
\begin{pmatrix}\rho_{1}(\xi_1-2\i,\xi_2)\\\rho_{2}(\xi_1-2\i,\xi_2)\\\rho_{3}(\xi_1-2\i,\xi_2)\end{pmatrix}=
L(\xi_1-\xi_2)\cdot\begin{pmatrix}\rho_{1}(\xi_1,\xi_2)\\\rho_{2}(\xi_1,\xi_2)\\\rho_{3}(\xi_1,\xi_2)\end{pmatrix}\ \text{.}
\end{equation}
Using the shorthand notation $\xi:=\xi_1-\xi_2$ the matrix $L(\xi)$ is given
by
\begin{equation}\label{Lmatrix}
L(\xi)=\begin{pmatrix}
\frac{\xi^2(20+\xi^2)}{(\xi-2\i)(\xi+2\i)(\xi-4\i)(\xi+4\i)} 
& \frac{4\i \xi(4-4\i  \xi+\xi^2)}{(\xi-2\i)(\xi+2\i)(\xi-4\i)(\xi+4\i)} &
\frac{4\i\xi}{(\xi+2\i)(\xi+4\i)} \\
 -\frac{16 \xi^2}{(\xi-2\i)(\xi+2\i)(\xi-4\i)(\xi+4\i)} 
& \frac{-4\i \xi (8-6\i
   \xi+\xi^2)}{(\xi-2\i)(\xi+2\i)(\xi-4\i)(\xi+4\i)} 
& \frac{\xi(\xi-2\i)}{(\xi+2\i)(\xi+4\i)} \\
 \frac{16(12+\xi^2)}{(\xi-2\i)(\xi+2\i)(\xi-4\i)(\xi+4\i)} 
& \frac{64-16\i \xi-4 \xi^2-8\i \xi^3+\xi^4}{(\xi-2\i)(\xi+2\i)(\xi-4\i)(\xi+4\i)} & \frac{-4\i(\xi-2\i)}{(\xi+2\i)(\xi+4\i)} 
\end{pmatrix}
\end{equation}
The functional equations for the three coefficients may be disentangled to 
some degree by the transformation
\begin{equation}\label{TrafoFGH}
\begin{pmatrix}F\\G\\H\end{pmatrix}(\xi_1,\xi_2)=
\begin{pmatrix}
9 & 3 & 3 \\
-16 & -{\xi}^2-8 & {\xi}^2+4 \\
-8 & -\frac 12{\xi}^2-6 & \frac 12{\xi}^2-6
\end{pmatrix}
\begin{pmatrix}\rho_{1}\\\rho_{2}\\\rho_{3}\end{pmatrix}(\xi_1,\xi_2).
\end{equation}
In fact, $F$ is fixed by the trace condition of the density operator
\begin{equation}\label{TraceF}
F(\xi_1,\xi_2)=9\rho_{1}(\xi_1,\xi_2)+3\rho_{2}(\xi_1,\xi_2)+3\rho_{3}(\xi_1,\xi_2)=\operatorname{tr}D^{[2]}(\xi_1,\xi_2)\equiv
1\ \text{.}
\end{equation}
Hence we find the simpler looking functional equations
\begin{equation}\label{BestimmungsgleichungenGH}
\begin{pmatrix}1\\G\\H\end{pmatrix}_{shift}=
\begin{pmatrix}1&0&0\\0& -\frac{\xi(\xi-6\i)}{(\xi-2\i)(\xi+4\i)} & 0 \\
-\frac{256i(\xi-\i)}{3(\xi+2\i)(\xi-2\i)^2(\xi+4\i)}& 
-\frac{\xi(\xi-6\i)(\xi^2-2\i\xi-4)}{(\xi-2\i)^2(\xi+2\i)(\xi+4\i)} 
& \frac{\xi^2(\xi-6\i)(\xi-4\i)}{(\xi-2\i)^2(\xi+2\i)(\xi+4\i)}\end{pmatrix}
\begin{pmatrix}1\\G\\H\end{pmatrix},
\end{equation}
where the functions $G,H$ on the left hand side are taken at shifted arguments
$(\xi_1-2\i,\xi_2)$, and those on the right hand side at $(\xi_1,\xi_2)$.
The inverse transformation is
\begin{equation}\label{UmkehrTrafoFGH}
\begin{pmatrix}\rho_{1}(\xi_1,\xi_2)\\\rho_{2}(\xi_1,\xi_2)\\\rho_{3}(\xi_1,\xi_2)\end{pmatrix}=
\begin{pmatrix} \frac{5\xi^2+36}{45(\xi^2+4)} &  -\frac{\xi^2}{30(\xi^2+4)} &  \frac{\xi^2+6}{15(\xi^2+4)} \\  -\frac{64}{45(\xi^2+4)} &  \frac{3\xi^2-20}{60(\xi^2+4)} &  -\frac{3\xi^2+28}{30(\xi^2+4)} \\ \frac{16}{45(\xi^2+4)} &  \frac{3\xi^2+20}{60(\xi^2+4)} &  -\frac{3\xi^2+8}{30(\xi^2+4)}
\end{pmatrix}
\begin{pmatrix}1\\G(\xi_1,\xi_2)\\H(\xi_1,\xi_2)\end{pmatrix}.
\end{equation}
Below, we show that the explicit and unique solution for the functions
$\rho_{j}(\xi_1,\xi_2)$, $j=1,2,3$, is
\begin{subequations}\label{Endergebnis2abc}
{\allowdisplaybreaks
\begin{align}
\rho_{1}(\xi_1,\xi_2)=&-\frac{1}{N(\xi_1)N(\xi_2)}
\frac 1{180 \xi^2 \left(4+\xi^2\right)}\biggl\{ -\frac94(5 \xi^2\left(4+\xi^2\right))
\nonumber\\
&\quad -\frac{3}{2}(5 \xi^2\left(4+\xi^2\right)-96) \left(
\omega(\xi_1-\i,\xi_1+\i)+\omega(\xi_2-\i,\xi_2+\i)\right)\nonumber\\
&\quad -(5 \xi^2\left(4+\xi^2\right)-96) \left( \omega(\xi_1-\i,\xi_1+\i)
\omega(\xi_2-\i,\xi_2+\i)\right) \nonumber\\
&\quad  -3\left(4+\xi^2\right)((\xi+2 \i) (\xi-6 \i) \omega
(\xi_1-\i,\xi_2+\i)+(\xi-2\i) (\xi+6 \i) \omega (\xi_1+\i,\xi_2-\i)) 
\nonumber\\
&\quad  +3 \xi^2 \left(16+\xi^2\right) (\omega (\xi_1-\i,\xi_2-\i)+\omega
(\xi_1+\i,\xi_2+\i)) \nonumber\\
&\quad +\left(6+\xi^2\right) \Bigl(\xi^2 \left(16+\xi^2\right) \omega
(\xi_1-\i,\xi_2-\i) \omega (\xi_1+\i,\xi_2+\i) \nonumber\\
&\quad-\left(4+\xi^2\right)^2 \omega (\xi_1-\i,\xi_2+\i) 
\omega (\xi_1+\i,\xi_2-\i)\Bigr) \biggr\}\ \text{,}
\\
\rho_{2}(\xi_1,\xi_2)=&-\frac{1}{N(\xi_1)N(\xi_2)}
\frac 1{180 \xi^2 \left(4+\xi^2\right)}\biggl\{ \nonumber\\
&\quad 12\left(5 \xi^2-28\right) \left(
\omega(\xi_1-\i,\xi_1+\i)+\omega(\xi_2-\i,\xi_2+\i)\right)\nonumber\\
&\quad + 8 \left(5 \xi^2-28\right) \left( \omega(\xi_1-i,\xi_1+i)
\omega(\xi_2-\i,\xi_2+\i)\right) \nonumber\\
&\quad  -3\left(4+\xi^2\right)((\xi+2 \i) (\xi+14 \i) \omega
(\xi_1-\i,\xi_2+\i)+(\xi-2 \i) (\xi-14 \i) \omega (\xi_1+\i,\xi_2-\i)) 
\nonumber\\
&\quad - 12 \xi^2\left(16+\xi^2\right) (\omega (\xi_1-\i,\xi_2-\i)+\omega
(\xi_1+\i,\xi_2+\i)) \nonumber\\
&\quad -\frac{1}{2} \left(28+3 \xi^2\right) \Bigl(\xi^2 \left(16+\xi^2\right)
\omega (\xi_1-\i,\xi_2-\i) \omega (\xi_1+\i,\xi_2+\i) \nonumber\\
&\quad -\left(4+\xi^2\right)^2 \omega (\xi_1-\i,\xi_2+\i) 
\omega (\xi_1+\i,\xi_2-\i)\Bigr) \biggr\}\ \text{,}
\\
\rho_{3}(\xi_1,\xi_2)=&-\frac{1}{N(\xi_1)N(\xi_2)}
\frac 1{180 \xi^2 \left(4+\xi^2\right)}\biggl\{\nonumber\\
&\quad -12\left(8+5 \xi^2\right) \left(
\omega(\xi_1-\i,\xi_1+\i)+\omega(\xi_2-\i,\xi_2+\i)\right) \nonumber\\
&\quad -8 \left(8+5 \xi^2\right) \left( \omega(\xi_1-\i,\xi_1+\i)
\omega(\xi_2-\i,\xi_2+\i)\right) \nonumber\\
&\quad  +12 \left(4+\xi^2\right) ((\xi-\i) (\xi+2 \i) 
\omega (\xi_1-\i,\xi_2+\i)+
(\xi-2 \i)(\xi+\i) \omega (\xi_1+\i,\xi_2-\i)) \nonumber\\
&\quad +3\xi^2\left(16+\xi^2\right) (\omega (\xi_1-\i,\xi_2-\i)+\omega
(\xi_1+\i,\xi_2+\i)) \nonumber\\
&\quad -\frac{1}{2} \left(8+3 \xi^2\right) \Bigl( \xi^2 \left(16+\xi^2\right)
\omega (\xi_1-\i,\xi_2-\i) \omega (\xi_1+\i,\xi_2+\i) \nonumber\\
&\quad -\left(4+\xi^2\right)^2 \omega (\xi_1-\i,\xi_2+\i) \omega (\xi_1+\i,\xi_2-\i) \Bigr)\biggr\}\ \text{.}
\end{align}}
\end{subequations}
Note that the function $N(\xi)$ appearing in the denominator is defined in
(\ref{defNo}).

\noindent
{\em Proof:} The symmetry of the function $\omega$
\begin{equation}
\omega (\xi_1,\xi_2)=\omega (\xi_2,\xi_1)\ ,
\end{equation} 
directly leads to 
the symmetry of the coefficients $\rho_{k}(\xi_1,\xi_2)$. 
The asymptotics of the function $\omega$
\begin{equation}
\lim_{\xi_{j}\to\infty}\omega (\xi_1,\xi_2)=0\qquad\hbox{for any\ }j=1,2,
\end{equation}
as obtained from \eqref{asymptotD}, for instance, entails 
the asymptotics of the functions $\rho_{k}(\xi_1,\xi_2)$
\begin{equation}
\lim_{\xi_{j}\to\infty}\rho_{k}(\xi_1,\xi_2)=\begin{cases}\frac19&\text{ for }k=1,\\0&\text{ else.}\end{cases}
\end{equation}
The functions $\rho_{k}(\xi_1,\xi_2)$ as solutions of the functional equations were constructed in the following way:
\eqref{Dreipunkt} is independent of the second spectral parameter, hence also the function 
\begin{multline}\label{defg}
g(\xi_1,\xi_2):=\frac{C}{N(\xi_1)N(\xi_2)}\:  \\  \Bigl\{\Omega(\xi_1+\i,\xi_2+\i)+\Omega(\xi_1-\i,\xi_2+\i) +\Omega(\xi_1+\i,\xi_2-\i)+\Omega(\xi_1-\i,\xi_2-\i)\\ - o(\xi_1-\xi_2-\i)-o(\xi_1-\xi_2+\i)\Bigr\}\:
\end{multline}
satisfies \eqref{Dreipunkt}, where $C$ is some constant. 
Then it is easy to show that the function
\begin{multline}\label{defh}
h(\xi_1,\xi_2):=\frac{C}{N(\xi_1)N(\xi_2)}\:\biggl\{ 
\Omega(\xi_1+\i,\xi_2+\i)\Omega(\xi_1-\i,\xi_2-\i)
-\Omega(\xi_1-\i,\xi_2+\i)\Omega(\xi_1+\i,\xi_2-\i) \\ +\Omega(\xi_1-\i,\xi_2+\i)o(\xi_1-\xi_2+\i)+ \Omega(\xi_1+\i,\xi_2-\i)o(\xi_1-\xi_2-\i)-o(\xi_1-\xi_2+\i)o(\xi_1-\xi_2-\i)\biggl\}\:
\end{multline}
satisfies
\begin{equation}
h(\xi_1,\xi_2)-h(\xi_1-2\i,\xi_2)=o(\xi_1-\xi_2-\i)g(\xi_1,\xi_2)
\end{equation}
Define
\begin{subequations}\label{Defgh}
\begin{align}
G(\xi_1,\xi_2)&:=\frac{1}{2\i}(\xi^2+16)(\xi^2+4)g(\xi_1,\xi_2)\ \text{ ,
}\\ H(\xi_1,\xi_2)&:=\frac{1}{(2\i)^2}(\xi^2+16)(\xi^2+4)^2 h(\xi_1,\xi_2)+
\frac{4}{3}\left(\frac{2\i}{\xi}\right)^2\ \text{,}
\end{align}
\end{subequations}
then these functions satisfy \eqref{BestimmungsgleichungenGH}. Applying the
inverse transformation \eqref{UmkehrTrafoFGH} the constant in \eqref{defg}
and \eqref{defh} is fixed by the analyticity requirement for the functions $\rho_{k}(\xi_1,\xi_2)$, leading to the
cancellation of spurious poles at $\xi_1\equiv\xi_2$ . Using the identity
\eqref{Lambdaratio} or \eqref{defNo} respectively leads to the results
\eqref{Endergebnis2abc}.\\

Finally, we consider the low and high temperature limits of
$\rho_{k}(\xi_1,\xi_2)$ in the homogeneous case $\xi_1=\xi_2=0$.  With the
results we easily calculate any nearest neighbour correlators like for instance
the internal energy $e=-T^2\frac{\partial}{\partial T}\left(f/T\right)$ by
\begin{equation}
\left.e\right|_{h=0}= -\frac{3J}{2}\left[2\rho_{1}(0,0)+3\rho_{3}(0,0)\right].
\end{equation}

\noindent
{\em Low temperature limit, zero field}

By use of (\ref{omegazerolimit}) we obtain 
\begin{equation}
\rho_{1}(0,0)=\frac{1}{2}-\frac{2}{45}\pi^2,\quad
\rho_{2}(0,0)=-\frac{19}{18}+\frac{14}{135}\pi^2,\quad
\rho_{3}(0,0)=-\frac{1}{9}+\frac{4}{135}\pi^2.
\end{equation}
Seemingly no zeta function values appear in these expressions. At first sight,
this looks qualitatively different from the case of the spin-1/2 Heisenberg
chain. Here, for the spin-1 chain, we find a $\pi^2$ expression instead of
zeta functions at odd integers. However, this can be rewritten in terms of
Riemann's zeta function at the special argument $\zeta(2)=\frac{\pi^2}{6}$.
We will see below that this pattern continues and for the 3-site density
matrix we obtain rational degree 3 polynomials in $\pi^2$ or equivalently
linear expressions in $\zeta(2)$, $\zeta(4)$ and $\zeta(6)$ with rational 
coefficients.\\
For the energy in the zero-temperature limit we simply get
\begin{equation}
\left.e\right|_{h=0,T=0}=-J,
\end{equation}
without any $\pi^2$ contributions. This result is in agreement with the
literature, cf. \cite{Takhtajan82}.\\

\noindent
{\em High temperature asymptotics, zero field}

By use of (\ref{omegahighlimit}) we obtain 
\begin{align}
\rho_{1}(0,0)&\simeq\frac{1}{9}-\frac{1}{108}\frac JT+\mathcal{O}(T^{-2}),\\
\rho_{2}(0,0)&\simeq-\frac{1}{36}\frac JT+\mathcal{O}(T^{-2}),\\
\rho_{3}(0,0)&\simeq\frac{1}{18}\frac JT+\mathcal{O}(T^{-2}),
\end{align}
and for the energy
\begin{equation}
\left.e\right|_{h=0}=-\frac{J}{3}-\frac{2}{9}\frac{J^2}T+\mathcal{O}(T^{-2})\ .
\end{equation}

\section{Computation of the 3-site density matrix}
\label{sec_3sites}
\begin{figure}[!h] 
\begin{center}
\includegraphics[width=10cm]{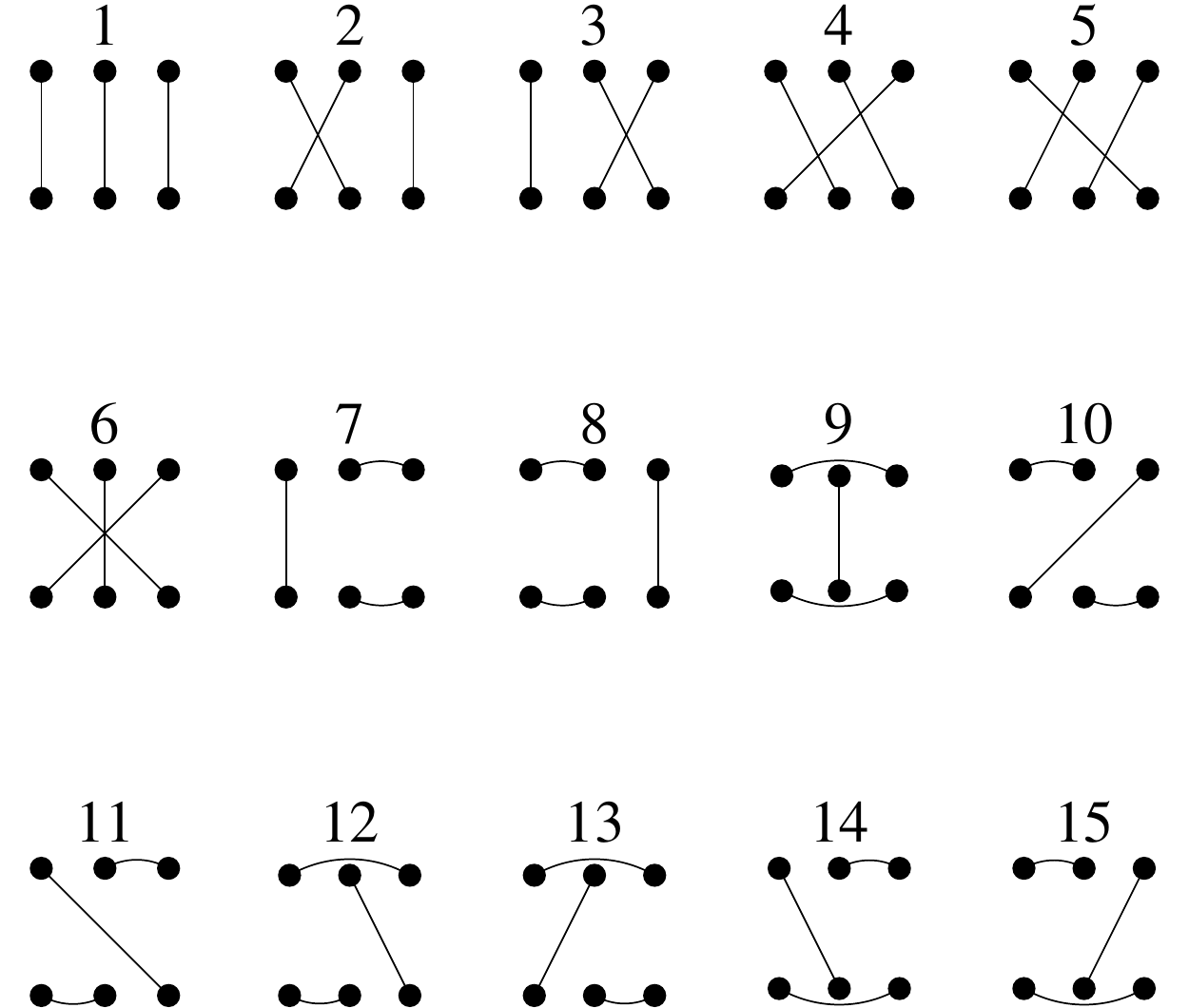}
\end{center}
\caption{}
 \label{p3}
\end{figure}

Similar to the 2-site case in the preceding section, the density operator of
a 3-site segment is presented in the form,
\begin{equation}
D^{[2]}(\xi_1,\xi_2,\xi_3) = \sum_{\alpha=1}^{15}
  \ntrho{\alpha}(\xi_1,\xi_2,\xi_3)  P_{\alpha}.
\end{equation}
The projectors $P_{\alpha}\,(1\le \alpha \le 15)$  are graphically represented in
Fig.~\ref{p3}.
The lines connecting upper and lower rows indicate Kronecker deltas.
The lines connecting nodes $a$ and $b$ ( $1\le a<b\le 3$ ) in the same row
represent matrix elements $ C(i_a, i_b) \delta_{i_a+i_b,0}$ such that 
\begin{equation}
C(i_a, i_b) =
\begin{cases}
 1&     (i_a,i_b) = (0,0) \\
 -1&     (i_a,i_b) = (1,-1),(-1,1).
\end{cases}
\end{equation}
This choice of projectors ensures the $SU(2)$ symmetry of the density operator,
\begin{equation}
G D^{[2]}(\xi_1, \xi_2,\xi_3) G^{-1} = D^{[2]}(\xi_1, \xi_2,\xi_3),
\qquad G\in {\rm SU(2)}.
\end{equation}
For later convenience, we also introduce renormalized coefficients
\begin{equation}
\wtrho{a}(\xi_1, \xi_2,\xi_3)=\prod_{j=1}^3 N(\xi_j) \ntrho{a}(\xi_1,\xi_2,\xi_3), 
\end{equation}
where $N(\xi_j)$ is defined in (\ref{defNo}).

The result of our calculations takes a compact form in the homogeneous limit
$\xi_1=\xi_2=\xi_3=0$ at $T=0$ (where $\ntrho{a}=8\wtrho{a}$):
\begin{align}
\wtrho{1}&=\frac{1879}{432}-\frac{3497}{1350}\pi^2+\frac{53}{135}\pi^4-\frac{11296}{637875}\pi^6\nonumber\\
\wtrho{2}&=
\wtrho{3}=-\frac{953}{2700}\pi^2+\frac{37}{675}\pi^4+\frac{1043}{2160}-\frac{1552}{637875}\pi^6\nonumber\\
\wtrho{4}&=
\wtrho{5}=\frac{983}{900}\pi^2-\frac{251}{144}+\frac{1592}{212625}\pi^6-\frac{1}{6}\pi^4\nonumber\\
\wtrho{6}&=\frac{197}{54}\pi^2-\frac{374}{675}\pi^4-\frac{13021}{2160}+\frac{3184}{127575}\pi^6\\
\wtrho{7}&=
\wtrho{8}=\frac{2189}{1350}\pi^2-\frac{166}{675}\pi^4-\frac{2917}{1080}+\frac{7072}{637875}\pi^6\nonumber\\
\wtrho{9}&=\frac{641}{225}\pi^2-\frac{292}{675}\pi^4-\frac{5119}{1080}+\frac{12448}{637875}\pi^6\nonumber\\
\wtrho{{10}}&=
\wtrho{{11}}=-\frac{253}{135}\pi^2+\frac{371}{120}-\frac{544}{42525}\pi^6+\frac{64}{225}\pi^4\nonumber\\
\wtrho{{12}}&=
\wtrho{{13}}=
\wtrho{{14}}=
\wtrho{{15}}=\frac{343}{360}-\frac{751}{1350}\pi^2+\frac{19}{225}\pi^4-\frac{272}{70875}\pi^6\nonumber
\end{align}
Although at the intermediate stages nontrivial numbers like Euler's $\gamma$ occur,
the final results are rational polynomials of degree 3 in $\pi^2$ and hence
are simply expressible by use of Riemann's zeta function at even integers
\begin{equation}
\zeta(2)=\frac{\pi^2}{6}, \quad \zeta(4)=\frac{\pi^4}{90}, \quad \zeta(6)=\frac{\pi^6}{945}.
\end{equation}
In the remainder of this section we describe the computational strategy.

Firstly, we note that not all of coefficients are independent but some of them are related,
reflecting the diagrammatic symmetry,
\begin{align}
\ntrho5(\xi_1, \xi_2, \xi_3)&=\ntrho4(-\xi_1, -\xi_2, -\xi_3)&
\ntrho{11}(\xi_1, \xi_2, \xi_3)&=\ntrho{10}(-\xi_1, -\xi_2, -\xi_3)   \nonumber \\
\ntrho{15}(\xi_1, \xi_2, \xi_3)&=\ntrho{12}(-\xi_1, -\xi_2, -\xi_3)&
\ntrho{14}(\xi_1, \xi_2, \xi_3)&=\ntrho{13}(-\xi_1, -\xi_2, -\xi_3)   \nonumber  \\
\ntrho3(\xi_1, \xi_2, \xi_3)&=\ntrho2(\xi_3, \xi_2, \xi_1)&
\ntrho8(\xi_1, \xi_2, \xi_3)&=\ntrho7(\xi_3, \xi_2, \xi_1)     \nonumber\\
\ntrho{14}(\xi_1, \xi_2, \xi_3)&=\ntrho{12}(\xi_3, \xi_2, \xi_1)&
\ntrho{15}(\xi_1, \xi_2, \xi_3)&=\ntrho{13}(\xi_3, \xi_2, \xi_1). \label{rhosymmetry}
\end{align}
The first four relations hold due to the up-down symmetry and the last four
ones due to the symmetry w.r.t. reflection
at the anti-diagonal line.
This reduces our task considerably.
The explicit expressions of $\ntrho{j}$ ($\wtrho{j} $) in
terms of $\omega$ functions are still very much involved.  Below we present
the simplest case $ \wtrho{1} $ for illustration.
Supplementary arguments are given in Appendix \ref{app_3site}.  The expression
for $ \wtrho{1} $ contains up to trilinear terms in $\omega$,
\begin{align*}
&\wtrho{1}(\xi_1,\xi_2,\xi_3)
= \frac{N(\xi_1)N(\xi_2)N(\xi_3)}{27} +
c^{(1)}_1 \omega(\xi_1^{-},\xi_2^{-})+c^{(1)}_2
   \omega(\xi_1^{-},\xi_1^{+})+c^{(1)}_3  \omega(\xi_1^{-},\xi_2^{+})\\
& +c^{(2)}_1 \omega(\xi_1^{-},\xi_1^{+})\omega(\xi_2^{-},\xi_3^{-}) 
  +c^{(2)}_2 \omega(\xi_1^{-},\xi_2^{-})\omega(\xi_2^{+},\xi_3^{-}) 
+c^{(2)}_3 \omega(\xi_1^{-},\xi_1^{+})   \omega(\xi_2^{-},\xi_3^{+})\\
& +c^{(2)}_4 \omega(\xi_1^{+},\xi_3^{-})\omega(\xi_2^{-},\xi_3^{+})
 +c^{(2)}_5 \omega(\xi_1^{-},\xi_2^{-})\omega(\xi_1^{+},\xi_3^{+})
 +c^{(2)}_6 \omega(\xi_2^{-},\xi_3^{+}) \omega(\xi_2^{+},\xi_3^{-})\\
   &+c^{(2)}_7 \omega(\xi_1^{-},\xi_1^{+})\omega(\xi_2^{-},\xi_2^{+})
 +c^{(2)}_8  \omega(\xi_2^{-},\xi_3^{-})\omega(\xi_2^{+},\xi_3^{+})
 +c^{(3)}_1 \omega(\xi_1^{-},\xi_1^{+}) \omega(\xi_2^{-},\xi_3^{+}) \omega(\xi_2^{+},\xi_3^{-})\\
 &+c^{(3)}_2\omega(\xi_1^{-},\xi_2^{+}) \omega(\xi_1^{+},\xi_3^{-}) \omega(\xi_2^{-},\xi_3^{+}) 
+c^{(3)}_3 \omega(\xi_1^{-},\xi_1^{+})\omega(\xi_2^{-},\xi_2^{+}) \omega(\xi_3^{-},\xi_3^{+})\\
& +c^{(3)}_4\omega(\xi_1^{-},\xi_2^{-}) \omega(\xi_1^{+},\xi_3^{-})
   \omega(\xi_2^{+},\xi_3^{+}) 
   +c^{(3)}_5 \omega(\xi_1^{-},\xi_1^{+})
   \omega(\xi_2^{-},\xi_3^{-}) \omega(\xi_2^{+},\xi_3^{+})\\
 & +\text{permutations and negation}.
   \end{align*}
We use the shorthand notation, $\xi^{\pm}_j:=\xi_j \pm \i$.  The last term
contains the distinct terms under permutations of $\{\xi_i\}$ and negation of
spectral parameters $\xi_i \rightarrow -\xi_i$.  We regard ``distinct'' after
applying the symmetry properties $\omega(\lambda_i ,\lambda_j)=\omega(\lambda_j ,\lambda_i)$ and
$\omega(\lambda_i ,\lambda_j)=\omega(-\lambda_i ,-\lambda_j)$.  The coefficients
$c^{(i)}_j$ are rational functions of $\xi_1, \xi_2,\xi_3$ and their explicit
forms are listed.
\begin{align*}
c^{(1)}_1 (\xi_1,\xi_2,\xi_3) &= -  \frac{\left(\xi_{12}^2+4^2\right) \left(7 \xi_{13}
   (\xi_{13}-2 \i) \xi_{23} (\xi_{23}-2 \i)+80
   \xi_{12}^2+2432\right)}{1680 (\xi_{12}-2 \i)
   (\xi_{12}+2 \i) \xi_{13} (\xi_{13}-2 \i)
 \xi_{23} (\xi_{23}-2 \i)}\\
c^{(1)}_2 (\xi_1,\xi_2,\xi_3) &=
   \frac{-7 (-\xi_{12} \xi_{23}+2)^2-7 (\xi_{13}\xi_{23}+2)^2
   -7 \xi_{12}\xi_{13}
   \left(\xi_{12}^2+\xi_{13}^2+\xi_{23}^2
   +8\right)-2056}{35 \xi_{12}^2 (\xi_{12}-2 \i)
   (\xi_{12}+2 \i) \xi_{13}^2 (\xi_{13}-2 \i)
   (\xi_{13}+2 \i)}\\
c^{(1)}_3 (\xi_1,\xi_2,\xi_3) &=
   \frac{(\xi_{12}+2 \i) (\xi_{12}-6 \i) \left(7
   (\xi_{13}\xi_{23}+2)^2+2 \xi_{12} (7
   \i \xi_{13} \xi_{23}+40 (\xi_{12}-4
   \i))+2084\right)}{1680 \xi_{12}^2 \xi_{13}
   (\xi_{13}-2 \i) \xi_{23} (\xi_{23}+2 \i)}\\
c^{(2)}_1 (\xi_1,\xi_2,\xi_3)& =   
 \frac{(\xi_{23}+4 \i) (-\xi_{23}+4 \i) p^{(2)}_1(\xi_1,\xi_2,\xi_3)}{12600
   \xi_{12}^2 (\xi_{12}-2 \i) (\xi_{12}+2 \i)
   \xi_{13}^2 (\xi_{13}-2 \i) (\xi_{13}+2 \i)
   (\xi_{23}-2 \i) (\xi_{23}+2 \i)}\\
c^{(2)}_2 (\xi_1,\xi_2,\xi_3) &=
  -\frac{i (\xi_{12}-4 \i) (\xi_{12}+4 \i)
   (\xi_{23}-4 \i) \left(-6e_2^2+832e_2+e_3^2-50
   \xi_{21}^2 \xi_{23}^2-1520 \xi_{21}
   \xi_{23}-27328\right)}{3150 \xi_{12}
   (\xi_{12}-2 \i) (\xi_{12}+2 \i) \xi_{13}
   (\xi_{13}-2 \i) (\xi_{13}+2 \i) \xi_{23}^2} \\
  c^{(2)}_3 (\xi_1,\xi_2,\xi_3) &=\frac{(\xi_{23}+2 \i)p^{(2)}_3( \xi_1,\xi_2,\xi_3   ) }{12600 \xi_{12}^2
   (\xi_{12}-2 \i) (\xi_{12}+2 \i) \xi_{13}^2
   (\xi_{13}-2 \i) (\xi_{13}+2 \i) \xi_{23}^2} \\
c^{(2)}_4 (\xi_1,\xi_2,\xi_3)& =   
\frac{ (\xi_{13}-4 \i) (\xi_{23}+4 \i)  p^{(2)}_4( \xi_1,\xi_2,\xi_3   ) }{3150 \i
   \xi_{12}^2 (\xi_{12}+2 \i) \xi_{13}^2
   \xi_{23}^2}  \\
c^{(2)}_5 (\xi_1,\xi_2,\xi_3) &=   
 \frac{(\xi_{12}-4 \i) (\xi_{12}+4 \i) (\xi_{13}-4
   \i) (\xi_{13}+4 \i) p^{(2)}_4( -\xi_2,-\xi_3,-\xi_1   )}{3150 \i \xi_{12}
   (\xi_{12}-2 \i) (\xi_{12}+2 \i) \xi_{13}
   (\xi_{13}-2 \i) (\xi_{13}+2 \i) \xi_{23}^2
   (\xi_{23}-2 \i)}  \\
c^{(2)}_6(\xi_1,\xi_2,\xi_3) &=\frac{3}{2}c^{(3)}_1(\xi_1,\xi_2,\xi_3)  \\
c^{(2)}_7(\xi_1,\xi_2,\xi_3) &=\frac{3}{2}c^{(3)}_3(\xi_1,\xi_2,\xi_3)  \\
c^{(2)}_8(\xi_1,\xi_2,\xi_3) &=\frac{3}{2}c^{(3)}_5(\xi_1,\xi_2,\xi_3)  \\
\end{align*}
\begin{align*}
c^{(3)}_1 (\xi_1,\xi_2,\xi_3)& =
\frac{p^{(3)}_1(\xi_1,\xi_2,\xi_3) (\xi_{23}-2 \i) (\xi_{23}+2 \i)}{37800
   \xi_{12}^2 (\xi_{12}-2 \i) (\xi_{12}+2 \i)
   \xi_{13}^2 (\xi_{13}-2 \i) (\xi_{13}+2 \i)
   \xi_{23}^2}\\
c^{(3)}_2 (\xi_1,\xi_2,\xi_3) &= -\frac{i \left(-6 e_2^2+432 e_2+e_3^2-5648\right)
   (\xi_{12}+4 \i) (\xi_{13}-4 \i) (\xi_{23}+4 \i)}{9450 \xi_{12}^2 \xi_{13}^2
   \xi_{23}^2}  \\
c^{(3)}_3 (\xi_1,\xi_2,\xi_3) & =
-\frac{4 \left(105 e_2^4-840 e_2^3+23632 e_2^2+420 e_2
   e_3^2+345856 e_2-4172 e_3^2+2891776\right)}{4725
 \xi_{12}^2 (\xi_{12}-2 \i) (\xi_{12}+2 \i)
   \xi_{13}^2 (\xi_{13}-2 \i) (\xi_{13}+2 \i)
   \xi_{23}^2 (\xi_{23}-2 \i) (\xi_{23}+2 \i)} \\
c^{(3)}_4 (\xi_1,\xi_2,\xi_3) &=\frac{i \left(-6 e_2^2+432 e_2+e_3^2-5648\right)
   (\xi_{12}-4 \i) (\xi_{12}+4 \i) (\xi_{13}-4 \i)
   (\xi_{23}-4 \i) (\xi_{23}+4 \i)}{9450
   \xi_{12} (\xi_{12}-2 \i) (\xi_{12}+2 \i)
   \xi_{13}^2 \xi_{23} (\xi_{23}-2 \i)
   (\xi_{23}+2 \i)} \\
c^{(3)}_5 (\xi_1,\xi_2,\xi_3) &=\frac{p^{(3)}_1(\xi_1,\xi_2,\xi_3) (\xi_{23}+4 \i) (-\xi_{23}+4 \i)}{37800
   \xi_{12}^2 (\xi_{12}-2 \i) (\xi_{12}+2 \i)
   \xi_{13}^2 (\xi_{13}-2 \i) (\xi_{13}+2 \i)
   (\xi_{23}-2 \i) (\xi_{23}+2 \i)}\\
\end{align*}
where $e_2$ and $e_3$ are elementary symmetric functions, defined as
coeffcients of $t^2$ and $t^3$ in the expansion of $(1+t \xi_{12})(1+t
\xi_{23})(1+t \xi_{31})$.  The nontrivial polynomials in the numerators are
defined by
\begin{align*}
  p^{(2)}_1(\xi_1,\xi_2,\xi_3) & =1748992+384 e_2^2+53248
   e_2+106496 \xi_{23}^2\\
   & +\xi_{12}\xi_{13}\left(-476 e_2
   \xi_{12}\xi_{13}+35 \xi_{12}^3
   \xi_{13}^3-196 \xi_{12}^2
   \xi_{13}^2+12720 \xi_{12}\xi_{13}-4800 \xi_{23}^2-136704\right)\\
 p^{(2)}_3( \xi_1,\xi_2,\xi_3   )   &=
 -5783552 \i  -606208 \xi_{23}
+384 e_2^2 \xi_{23}-2304 \i e_2^2\\
&+\xi_{12}^2
   \xi_{13}^2 (\xi_{23}-6 \i) \left(-476 e_2+35
   \xi_{12}^2 \xi_{13}^2-196 \xi_{12}
   \xi_{13}\right)
-9728 e_2 \xi_{23}+232448 \i e_2\\
    &  -80 \xi_{12}^2 \xi_{13}^2 (37
   \xi_{23}+562 \i)-320 \xi_{12} \xi_{13}
   \left(15 \xi_{23}^3-90 \i \xi_{23}^2-76
   \xi_{23}-2888 \i\right) \\
p^{(2)}_4( \xi_1,\xi_2,\xi_3   )  & =
22592-3280 \i \xi_{12}- 6e_2^2-88 e_2-60 \i \xi_{12}
   \left(\xi_{13}^2+\xi_{23}^2\right) \\
&  +\xi_{31} \xi_{32} \Bigl(1640+\left(\xi_{21}^2+10 \i
   \xi_{12}+30\right) \xi_{31} \xi_{32}\Bigr) \\
     p^{(3)}_1( \xi_1,\xi_2,\xi_3   ) &=5783552  -384
   \left(e_2^3+3 e_3^2\right)+e_2^2 \left(140
   \xi_{12}^2 \xi_{13}^2+33792\right)+5184 e_2
   \xi_{12} \xi_{13} \xi_{23}^2-803840 e_2\\
   &+\xi_{12} \xi_{13}
   \Bigl(\xi_{12} \xi_{13} \left(35 e_3^2+336
   \xi_{23}^4+13304 \xi_{23}^2+13472\right) \\
   &+70
   \xi_{12}^3 \xi_{13}^3+1296 \xi_{12}^2 \xi_{13}^2-169728\xi_{23}^2-1495552\Bigr).
\end{align*}
The above result is checked against direct computations (of some density
matrix elements ) for fixed Trotter numbers.

\section{Conclusion}

The present report offers a novel way to approach the exact evaluation of
correlation functions of higher spin $su(2)$ chains in a quantitative manner.
It utilizes the discrete functional relations of qKZ equation type, as well as the
direct fusion procedure.
Explicit results for shorter segments ($m=2,3$) clearly show an improvement
of the present approach compared to the former formulation in \cite{GSS10}.  The
correlation functions are not given by complex contour integrals but they are
given in simple factorized forms, using rational functions of inhomogeneities
and only one nontrivial ingredient $\omega$.
One of the direct consequences of such nice expressions is exemplified in Section \ref{sec_2sites},
and Section \ref{sec_3sites}:
correlation functions of the $S=1$ chain contain only Riemann's zeta function
with even integer arguments.
This complements the famous conjecture in \cite{BoKo01}
that correlation functions of the $S=\frac{1}{2}$ chain are given in terms of zeta
functions with odd integer arguments.

Clearly many subjects are still left open.  The appearance of the zeta
function with odd integer arguments for the $S=\frac{1}{2}$ chain is naturally
explained via the explicit construction of the solution to the qKZ equation using
$q$-oscillators.  We believe that an analogous reasoning is necessary and possible
for the higher spin case.

It becomes clearer that bulk quantities (e.g.~the specific heat) of some novel
materials, e.g., spin-ladders and spin-nanotubes are described by models with
higher rank Lie symmetry.  Their quantum correlations are definitely the next
promising goal of our understanding of these materials.  Therefore the
extension of the present approach to the su($n$) case is highly desirable.

We hope to come back to these issues in the near future.
\\[1ex]
{\bf Acknowledgment.} We would like to thank B.~Aufgebauer, H.~Boos, F.~G\"ohmann,
M.~Jimbo, A.~Kuniba, T. Miwa for
stimulating discussions. AK is grateful to Shizuoka University for
hospitality. His work was supported by JSPS.
JS is supported by a Grant-in-Aid for Scientific Research No.\ 20540370.


{\appendix
\Appendix{Auxiliary functions for spin 1}
\label{app:auxfun}
It is sometimes more convenient to deal with polynomials rather than
with rational functions. For this reason a different normalization of
the elementary $R$-matrix was used in \cite{Suzuki99}. This leads to
differently normalized transfer matrix eigenvalues. In order to simplify
the comparison with \cite{Suzuki99} we define the functions
\begin{subequations}
\begin{align}\label{lambdapol1}
      \La_1 (\la) =
       &\phi_- (\la - 3 \i) \phi_+ (\la + 3 \i) 
       \vphi_- (\la - 2 \i) \vphi_+ (\la + 2 \i) 
\cdot\La^{[1]} (\la) \epc \\[1ex]
\label{lambdapol2}
      \La_2 (\la) =
       &\phi_- (\la - 4 \i) \phi_+ (\la + 2 \i)
       \phi_- (\la - 2 \i) \phi_+ (\la + 4 \i)\cdot\nonumber\\
&       \vphi_- (\la - 3 \i) \vphi_+ (\la + \i)
       \vphi_- (\la - \i) \vphi_+ (\la + 3 \i)
       \cdot\La^{[2]} (\la) \epp
\end{align}
\end{subequations}
Then, following \cite{Suzuki99}, we introduce
\begin{subequations}
\begin{align}
     & \la_1 (\la) = \re^{- \frac{2h}{T}} \phi(\la - \i)
                     \phi(\la -3 \i) \vphi(\la) \vphi(\la-2\i)
                     \frac{q(\la + 3 \i)}{q(\la - \i)} \epc \label{lamb1}\\[1ex]
     & \la_2 (\la) = \phi(\la + \i) \phi(\la-\i) \vphi^2(\la)
                     \frac{q(\la - 3 \i) q(\la + 3 \i)}
                          {q(\la - \i) q(\la + \i)} \epc \\[1ex]
     & \la_3 (\la) = \re^{\frac{2h}{T}} \phi(\la+3\i) \phi(\la + \i)
                     \vphi(\la + 2 \i) \vphi(\la)
                     \frac{q(\la - 3 \i)}{q(\la + \i)} \epp\label{lamb3}
\end{align}
\end{subequations}
where
\begin{equation} \label{defphi}
\phi(\la):=\phi_+(\la+\i)\phi_-(\la-\i)\ , \ 
\vphi(\la):=\vphi_+(\la+\i)\vphi_-(\la-\i) .
\end{equation}
It follows that
\begin{equation}
     \La_2 (\la) = \la_1 (\la) + \la_2 (\la) + \la_3 (\la) \epp
\end{equation}
The basic auxiliary functions for spin 1 are
\begin{equation}
     \fb (\la) = \frac{\la_1 (\la + \i) + \la_2 (\la + \i)}
                      {\la_3 (\la + \i)} \epc \qd
     \fbq (\la) = \frac{\la_2 (\la - \i) + \la_3 (\la - \i)}
                       {\la_1 (\la - \i)} \epc\label{auxbbq}
\end{equation}
with corresponding capital functions
\begin{equation}
     \Bf (\la) = 1 + \fb (\la) \epc \qd \Bfq (\la) = 1 + \fbq (\la) \epp
\end{equation}

In \cite{Suzuki99} the nonlinear integral equations (\ref{nlie}) were
derived from a set of functional equations satisfied by the functions
$\fb, \fbq, \Bf, \Bfq$ together with
\begin{equation}\label{yY}
     y(\la)  = \frac{\La_2 (\la)}{\phi(\la - 3 \i) \phi(\la + 3 \i)
                     \vphi(\la - 2 \i) \vphi(\la + 2 \i)} \epc \qd
     Y(\la) = 1 + y(\la) \epp
\end{equation}
In appendix \ref{app:contnlietonlie} we present a derivation
of integral equations by use of algebraic relations exposed below.
\begin{subequations}
\begin{align}
     & \fb (\la) = \re^{-\frac{3h}{T}} \Lambda_1(\la)\frac{\phi(\la)}
                     {\phi(\la +4 \i)\phi(\la +2 \i) \vphi(\la+3\i)}
                     \frac{q(\la + 4 \i)}{q(\la -2 \i)} \epc \label{fb}\\[1ex]
     & \fbq (\la) = \re^{\frac{3h}{T}} \Lambda_1(\la)\frac{\phi(\la)}
                     {\phi(\la -4 \i)\phi(\la -2 \i) \vphi(\la-3\i)}
                     \frac{q(\la - 4 \i)}{q(\la +2 \i)} \epc \label{fbq}\\[1ex]
     & \Lambda_2(\la)=\Bf(\la-\i)\lambda_3(\la)=\Bfq(\la+\i)\lambda_1(\la)\epc
                     \label{Lambda2}\\[1ex]
     & Y(\la)  = \frac{\La_1 (\la-\i)\La_1 (\la+\i)}{\phi(\la - 3 \i) \phi(\la + 3 \i) \vphi(\la - 2 \i) \vphi(\la + 2 \i)} \epp\label{Y}
\end{align}
\end{subequations}

Note that from (\ref{yY}) and (\ref{Lambda2}) we obtain
\begin{equation}
1+y(\la)^{-1}=\frac{Y(\la)}{y(\la)}=\frac{\La_1 (\la+\i)\La_1 (\la-\i)}{\La_2(\la)}
\end{equation}

\Appendix{NLIE with straight contour integrations}
\label{app:contnlietonlie}\noindent

To proceed further, it is convenient to consider equations in Fourier
space. For a smooth function $f(\la)$ we define 
\begin{equation}
     \hat{f}(k) =
        \int_{-\infty}^{\infty} \frac{\rd \la}{2\pi}
	{\rm e}^{\i k\la} f(\la) \epc \qd
     {{\rm dlf}}(k) = \int_{-\infty}^{\infty}
	\frac{\rd \la}{2\pi} {\rm e}^{\i k\la}
        \Bigl(\frac{d}{d\la} \log f(\la) \Bigr) \epp
\end{equation}
Equations (\ref{Lambda2}), (\ref{lamb1}), (\ref{lamb3}) and (\ref{Y}), after 
taking the logarithmic derivative and the Fourier transform, read 
\begin{subequations}
\begin{align}
{{\rm dl\Lambda_2}}&={\rm e}^{-k} {{\rm dlB}}+{{\rm dl\lambda_3}}
={\rm e}^{k} {{\rm dl\bar B}}+{{\rm dl\lambda_1}},\label{dlL2}\\
{{\rm dl\lambda_1}}&=\left({\rm e}^{-k}+{\rm e}^{-3k}\right){{\rm dl\phi_-}}
+{{\rm dl\vphi_0}}+{\rm e}^{-2k}{{\rm dl\vphi_-}}
+{\rm e}^{3k}{{\rm dlq_+}}-{\rm e}^{-k}{{\rm dlq_-}},\label{dllam1}\\
{{\rm dl\lambda_3}}&=\left({\rm e}^{k}+{\rm e}^{3k}\right){{\rm dl\phi_+}}
+{{\rm dl\vphi_0}}+{\rm e}^{2k}{{\rm dl\vphi_+}}
+{\rm e}^{-3k}{{\rm dlq_-}}-{\rm e}^{k}{{\rm dlq_+}},\label{dllam3}\\
{{\rm dlY}}
&=\left({\rm e}^{k}+{\rm e}^{-k}\right) {{\rm dl\Lambda_1}}
-{\rm e}^{-3k}{{\rm dl\phi_-}}-{\rm e}^{3k}{{\rm dl\phi_+}}
-{\rm e}^{-2k}{{\rm dl\vphi_-}}-{\rm e}^{2k}{{\rm dl\vphi_+}}.\label{dlY}
\end{align}
\end{subequations}
These equations simplify in different ways depending on the sign of $k$.

\subsection*{$k>0$.}
Here ${{\rm dl\phi_+}}={{\rm dl\vphi_+}}={{\rm dlq_+}}=0$, hence from 
(\ref{dlL2})-(\ref{dllam3})
\begin{subequations}
\begin{align}
\nonumber
&{{\rm dl\lambda_3}}-{{\rm dl\lambda_1}}=\\
&-\left({\rm e}^{-k}+{\rm e}^{-3k}\right){{\rm dl\phi_-}}
-{\rm e}^{-2k}{{\rm dl\vphi_-}}
+\left({\rm e}^{-k}+{\rm e}^{-3k}\right){{\rm dlq_-}}
=-{\rm e}^{-k} {{\rm dlB}}+{\rm e}^{k} {{\rm dl\bar B}},
\end{align}
\end{subequations}
from which we can express ${{\rm dlq_-}}$ in terms of 
${{\rm dl\phi_-}}$, ${{\rm dl\vphi_-}}$, ${{\rm dlB}}$ and ${{\rm dl\bar
    B}}$. From (\ref{dlY}) we obtain ${{\rm dl\Lambda_1}}$ in terms of
${{\rm dl\phi_-}}$, ${{\rm dl\vphi_-}}$ and ${{\rm dlY}}$. These expressions
are then inserted into the Fourier transforms of the logarithmic derivatives
of (\ref{fb}) and (\ref{Y}) yielding
\begin{subequations}
\begin{align}
{{\rm dlb}}&={{\rm dl\phi_0}}+\frac 1{{\rm e}^{k}+{\rm e}^{-k}}
\left(-{\rm e}^{-k}{{\rm dl\phi_-}}+{{\rm dlY}}+
{\rm e}^{-k} {{\rm dlB}}-{\rm e}^{k} {{\rm dl\bar B}}\right),\\
{{\rm dly}}&={{\rm dl\vphi_0}}+\frac 1{{\rm e}^{k}+{\rm e}^{-k}}
\left(-{\rm e}^{-k}{{\rm dl\vphi_-}}+ 
{{\rm dlB}}+{{\rm dl\bar B}}\right).
\end{align}
\end{subequations}
The explicit $\phi$ and $\vphi$ terms yield (note that ${{\rm dl\phi_-}}$
refers to the Fourier transform of the logarithmic derivative of $\phi(\la)$
in the lower half plane, it is not the Fourier transform of $\phi_-(\la)$):
\begin{subequations}
\begin{align}
{{\rm dl\phi_0}}-\frac {{\rm e}^{-k}}{{\rm e}^{k}+{\rm e}^{-k}}
{{\rm dl\phi_-}}&=-\i \frac N2\frac{
{\rm e}^{uk}-{\rm e}^{-uk}}{{\rm e}^{k}+{\rm e}^{-k}},\\
{{\rm dl\vphi_0}}-\frac {{\rm e}^{-k}}{{\rm e}^{k}+{\rm e}^{-k}}
{{\rm dl\vphi_-}}&=-\i\frac{1}{{\rm e}^{k}+{\rm
    e}^{-k}}{\rm e}^{\i\mu k}\left(1-{\rm e}^{\i\delta k}\right).
\end{align}
\end{subequations}

\subsection*{$k<0$.}
Here ${{\rm dl\phi_-}}={{\rm dl\vphi_-}}={{\rm dlq_-}}=0$, and a 
derivation similar to above yields
\begin{subequations}
\begin{align}
{{\rm dlb}}&={{\rm dl\phi_0}}+\frac 1{{\rm e}^{k}+{\rm e}^{-k}}
\left(-{\rm e}^{-k}{{\rm dl\phi_+}}+{{\rm dlY}}+
{\rm e}^{-k} {{\rm dlB}}-{\rm e}^{k} {{\rm dl\bar B}}\right),\\
{{\rm dly}}&={{\rm dl\vphi_0}}+\frac 1{{\rm e}^{k}+{\rm e}^{-k}}
\left(-{\rm e}^{-k}{{\rm dl\vphi_+}}+ 
{{\rm dlB}}+{{\rm dl\bar B}}\right),
\end{align}
\end{subequations}
with
\begin{subequations}
\begin{align}
{{\rm dl\phi_0}}-\frac {{\rm e}^{-k}}{{\rm e}^{k}+{\rm e}^{-k}}
{{\rm dl\phi_+}}&=-\i \frac N2\frac{
{\rm e}^{uk}-{\rm e}^{-uk}}{{\rm e}^{k}+{\rm e}^{-k}},\\
{{\rm dl\vphi_0}}-\frac {{\rm e}^{-k}}{{\rm e}^{k}+{\rm e}^{-k}}
{{\rm dl\vphi_+}}&=-\i\frac{1}{{\rm e}^{k}+{\rm
    e}^{-k}}{\rm e}^{\i\mu k}\left(1-{\rm e}^{\i\delta k}\right).
\end{align}
\end{subequations}

These results and an expression for ${{\rm dl\bar b}}$ similar to that for
${{\rm dlb}}$ are Fourier transformed, yielding integral equations for the
derivatives of the logarithms of the auxiliary functions $\fb$, $\fbq$ and $y$
in terms of $\Bf$, $\Bfq$ and $Y$. Integrating these equations and fixing the
integration constants from the known asymptotics completes our derivation of
the NLIE.\\

We are interested in $\La_1$ for which we derive from (\ref{dlY})
\begin{equation}
{{\rm dl\Lambda_1}}=\frac 1{{\rm e}^{k}+{\rm e}^{-k}}
\left({{\rm dlY}}+{\rm e}^{-3|k|}{{\rm dl\phi_\mp}}+
{\rm e}^{-2|k|}{{\rm dl\vphi_\mp}}\right).\label{dlL1}
\end{equation}

\Appendix{A simple nearest neighbour correlator}
\label{app:simplcorr}\noindent
We are going to relate the derivative of the eigenvalue
$\frac{\partial}{\partial\delta}
\ln\Lambda^{[1]}(\la)\Big|_{\delta=0}$
of the spin-1/2 column-to-column transfer matrix
$t^{\left[1\right]}(\la;\mu)$ with spectral parameter $\la$
of the spin-1 system modified by two additional horizontal spin-1/2 lines with
spectral parameters $\mu+\i$ and $\mu-\i+\delta$ to a matrix element of the
density operator $D^{[1]}(\la,\mu)$ of the unmodified spin-1 system for
two neighbouring spin-1/2 spaces with spectral parameters $\la, \mu$.

We follow \cite{AuKl12}. Be $\left\langle \Psi\right|$ and
$\left|\Psi\right\rangle$ normalized left and right eigenstates of
${t^{[1]}}(\la;\mu)$ with eigenvalue
${\Lambda^{[1]}}(x;\mu)$. We then have
{\allowdisplaybreaks
\begin{multline}
\frac{\partial}{\partial\delta}\left.\ln\left\{{\Lambda^{[1]}}(\la;\mu)\right\}\right|_{\delta=0}=
\frac{\partial}{\partial\delta}\left.\ln\left\{{\left\langle
  \Psi\right|
t^{\left[1\right]}}(\la;\mu)
\left|\Psi\right\rangle\right\}\right|_{\delta=0}=\left.\frac{\left\langle
  \Psi\right|\frac{\partial}{\partial\delta}{t^{\left[1\right]}}(\la;\mu)
  \left|\Psi\right\rangle}{\left\langle \Psi\right|
  {t^{\left[1\right]}}(\la;\mu)
  \left|\Psi\right\rangle}\right|_{\delta=0} ,\label{der_epsilon}
\end{multline}} 
where for the last equality we used that $\left\langle \Psi
|\Psi\right\rangle=1$ for all $\delta$. Next we use the factorization of 
$\left\langle \Psi\right|$ and $\left|\Psi\right\rangle$ at $\delta=0$ into
eigenstates of the original spin-1 system (without the pair of additional
horizontal spin-1/2 lines) times a singlet of two spin-1/2 objects. By use of 
the Yang-Baxter algebra and unitarity we find
\begin{figure}[h]
\centering
\resizebox{!}{4cm}{\input{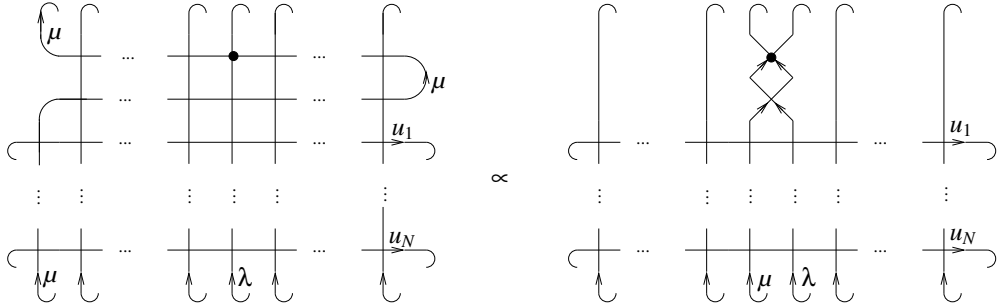}}
\caption{Graphical illustration of the last equation in (\ref{der_epsilon}).}
\label{fig:omega}
\end{figure}

\begin{equation}\label{TransfermatrixAbleitung}
\frac{\partial}{\partial\delta}\left.\ln\left\{{\Lambda^{[1]}}(\la;\mu)\right\}\right|_{\delta=0}=\frac{\operatorname{tr}\left\{D^{[1]}(\la,\mu)R(\la-\i,\mu-\i)\left.\frac{\partial}{\partial\delta}R{}(\mu+\delta-\i,\la-\i)\right|_{\delta=0}\right\}}{\operatorname{tr}\left\{D^{[1]}(\la,\mu)R(\la-\i,\mu-\i)R(\mu-\i,\la-\i)\right\}}.
\end{equation}
Using the normalization of the $\mathcal{R}$ matrices used in the
definition of ${t^{\left[1\right]}}(\la;\mu)$ we find unitarity in
the following form
\begin{equation}
\left.R(\la-\i,\mu-\i)R(\mu+\delta-\i,\la-\i)\right|_{\delta=0}=\mathds{1}
\end{equation}
and the derivative
\begin{equation}
R(\la-\i,\mu-\i)\left.\frac{\partial}{\partial\delta}R(\mu+\delta-\i,\la-\i)\right|_{\delta=0}=\frac{2\i}{(\la-\mu)^2+4)}\:\left[\mathds{1}-P^{[1]}\right].
\end{equation}
Upon introducing the function $\Omega(\la,\mu)$ by
\begin{equation}\label{Omegahalb}
\Omega(\la,\mu):=\frac{2\i}{(\la-\mu)^2+4)}\operatorname{tr}\left\{D^{[1]}(\la,\mu)\:P^{[1]}\right\}\ \text{,}
\end{equation}
we find (\ref{omegaTransfermatrix}) and the equivalent
(\ref{Integraldarstellungomega}).\\

\Appendix{More on 3-site density matrix}
\label{app_3site}
In Section \ref{sec_3sites}, the coefficients $\wtrho{j}$ were
introduced and only the explicit form of $\wtrho{1}$ was given.
Here we comment on the other ones.

Although their explicit forms in terms of $\omega$ are very involved as
already commented, the expressions in terms of particular 6-site density
matrix elements of spin $\frac{1}{2}$ chain are still manageable.

Let us define three elements
\begin{equation*}
P_6=\big(D^{[1]}\big)^{++++++}_{++++++},\quad
F_6=\big(D^{[1]}\big)^{+++++-}_{-+++++},\quad
G_6=\big(D^{[1]}\big)^{++++--}_{--++++},
\end{equation*}
where the spin-1/2 density matrix $D^{[1]}$ depends on the six arguments
$(\xi^+_1,\xi^-_1,\xi^+_2,\xi^-_2,\xi^+_3,\xi^-_3)$ where $\xi^{\pm}_j =\xi_j
\pm \i$ for $j=1,2,3$.  Then we find explicitly
\begin{footnotesize}
\begin{align*}
&\wtrho{1}(\xi_1,\xi_2,\xi_3)=
\frac{1}{27} {\cal N}(\xi_1,\xi_2,\xi_3)+\frac{14}{9} G_5(-\xi^{-}_1, -\xi^{-}_2, -\xi^{+}_2,-\xi^{-}_3, -\xi^{+}_3) \\
&+
\frac{14(2 \i+\xi_1-\xi_2)}{9 (\xi_1-\xi_2)} G_6(-\xi^{+}_1,-\xi^{-}_2, -\xi^{-}_1, -\xi^{+}_2,-\xi^{-}_3, -\xi^{+}_3)
+
\frac{2 (-4\i+\xi_2-\xi_3) }{3 (-2 \i+\xi_2-\xi_3)}G_6(-\xi^{-}_1, -\xi^{+}_1,-\xi^{+}_2,-\xi^{+}_3,-\xi^{-}_2, -\xi^{-}_3)  \\
&-
2(56 \i-28 \xi_1-8 \i \xi_1^2-4 \i \xi_1 \xi_2+\xi_1^2 \xi_2+12 \i \xi_2^2-2 \xi_1 \xi_2^2+\xi_2^3+28 \xi_3 +20
i \xi_1 \xi_3 -\xi_1^2 \xi_3 \\
&
-20 \i \xi_2 \xi_3+2 \xi_1 \xi_2 \xi_3-\xi_2^2 \xi_3   )
\times 
\frac{G_6(-\xi^{-}_1, -\xi^{+}_1,-\xi^{-}_2, -\xi^{+}_2,-\xi^{-}_3, -\xi^{+}_3)}
{9(\xi_1-\xi_2) (2 \i+\xi_1-\xi_2) (-2 \i+\xi_2-\xi_3)}
\\
& -\frac{4 (4 \i+\xi_1-\xi_2)}{2
i+\xi_1-\xi_2} G_6(-\xi^{-}_1, -\xi^{-}_2, -\xi^{+}_1,-\xi^{+}_2,-\xi^{-}_3, -\xi^{+}_3) 
-\frac{2 (4 \i+\xi_2-\xi_3) (-6 \i+\xi_2-\xi_3) }{3
(-2 \i+\xi_2-\xi_3) (2 \i+\xi_2-\xi_3)}G_6(\xi^{-}_1,\xi^{+}_1,\xi^{-}_2,\xi^{-}_3,\xi^{+}_2,\xi^{+}_3)\\
&-\frac{(2 \i+\xi_2-\xi_3) (4 \i+\xi_2-\xi_3) }{3
(\xi_2-\xi_3) (-2 \i+\xi_2-\xi_3)}G_6(\xi^{-}_1,\xi^{+}_1,\xi^{-}_3,\xi^{+}_3,\xi^{-}_2,\xi^{+}_2)
-\frac{2 (-4 \i+\xi_1-\xi_2) (6 \i+\xi_1-\xi_2) }{3
(-2 \i+\xi_1-\xi_2) (2 \i+\xi_1-\xi_2)}G_6(\xi^{+}_1,\xi^{+}_2,\xi^{-}_1,\xi^{-}_2,\xi^{-}_3,\xi^{+}_3)\\
&-\frac{(-2 \i+\xi_1-\xi_2) (-4 \i+\xi_1-\xi_2) }{3
(\xi_1-\xi_2) (2 \i+\xi_1-\xi_2)}G_6(\xi^{-}_2,\xi^{+}_2,\xi^{-}_1,\xi^{+}_1,\xi^{-}_3,\xi^{+}_3)\\
&- ( 48 \i \xi_1-24 \xi_1^2-96 \i \xi_2+4 \xi_1 \xi_2-10 \i \xi_1^2 \xi_2-4 \xi_2^2+30
i \xi_1 \xi_2^2+\xi_1^2 \xi_2^2-20 \i \xi_2^3-2 \xi_1 \xi_2^3+\xi_2^4+48 \i \xi_3\\
&+44 \xi_1 \xi_3+10 \i \xi_1^2
\xi_3+4 \xi_2 \xi_3-40 \i \xi_1 \xi_2 \xi_3-2 \xi_1^2 \xi_2 \xi_3+30 \i \xi_2^2 \xi_3+4 \xi_1 \xi_2^2
\xi_3-2 \xi_2^3 \xi_3-24 \xi_3^2+10 \i \xi_1 \xi_3^2\\
&+\xi_1^2 \xi_3^2-10 \i \xi_2 \xi_3^2-2 \xi_1 \xi_2
\xi_3^2+\xi_2^2 \xi_3^2 )
 \frac{G_6(\xi^{-}_1,\xi^{+}_1,\xi^{-}_2,\xi^{+}_2,\xi^{-}_3,\xi^{+}_3)}{9 (\xi_1-\xi_2)
(-2 \i+\xi_1-\xi_2) (\xi_2-\xi_3) (2 \i+\xi_2-\xi_3)}\\
\displaybreak[2]       
%
\\
&\wtrho{2}(\xi_1,\xi_2,\xi_3)=-\frac{1}{27} {\cal N}(\xi_1,\xi_2,\xi_3)
-\frac{14}{9} G_5(-\xi^{-}_1, -\xi^{-}_2, -\xi^{+}_2,-\xi^{-}_3, -\xi^{+}_3) \\
&-\frac{2 (4 \i+\xi_1-\xi_2) }{\xi_1-\xi_2}F_6(\xi^{+}_1,\xi^{-}_1,\xi^{-}_2,\xi^{+}_2,\xi^{-}_3,\xi^{+}_3)
-\frac{2
(-4 \i+\xi_1-\xi_2)}{\xi_1-\xi_2} F_6(\xi^{+}_2,\xi^{-}_1,\xi^{+}_1,\xi^{-}_2,\xi^{-}_3,\xi^{+}_3) \\
&-\frac{14
(2 \i+\xi_1-\xi_2)}{9 (\xi_1-\xi_2)}   G_6(-\xi^{+}_1,-\xi^{-}_2, -\xi^{-}_1, -\xi^{+}_2,-\xi^{-}_3, -\xi^{+}_3)
-\frac{2 (-4
i+\xi_2-\xi_3)}{3 (-2 \i+\xi_2-\xi_3)} G_6(-\xi^{-}_1, -\xi^{+}_1,-\xi^{+}_2,-\xi^{+}_3,-\xi^{-}_2, -\xi^{-}_3)  \\
&+2(56 \i-28 \xi_1-8 \i \xi_1^2-4 \i \xi_1 \xi_2+\xi_1^2 \xi_2+12 \i \xi_2^2-2 \xi_1 \xi_2^2+\xi_2^3+28 \xi_3+20
i \xi_1 \xi_3-\xi_1^2 \xi_3-20 \i \xi_2 \xi_3 \\
&+2 \xi_1 \xi_2 \xi_3-\xi_2^2 \xi_3 ) \frac{G_6(-\xi^{-}_1, -\xi^{+}_1,-\xi^{-}_2, -\xi^{+}_2,-\xi^{-}_3, -\xi^{+}_3)}{9
(\xi_1-\xi_2) (2 \i+\xi_1-\xi_2) (-2 \i+\xi_2-\xi_3)}\\
&+\frac{4 (4 \i+\xi_1-\xi_2)}{2
i+\xi_1-\xi_2} G_6(-\xi^{-}_1, -\xi^{-}_2, -\xi^{+}_1,-\xi^{+}_2,-\xi^{-}_3, -\xi^{+}_3) \\
&+(48 \i \xi_1-24 \xi_1^2-96 \i \xi_2+4 \xi_1 \xi_2-10 \i \xi_1^2 \xi_2-4 \xi_2^2+30
i \xi_1 \xi_2^2+\xi_1^2 \xi_2^2-20 \i \xi_2^3-2 \xi_1 \xi_2^3+\xi_2^4\\
&+48 \i \xi_3+44 \xi_1 \xi_3+10 \i \xi_1^2
\xi_3+4 \xi_2 \xi_3-40 \i \xi_1 \xi_2 \xi_3-2 \xi_1^2 \xi_2 \xi_3+30 \i \xi_2^2 \xi_3+4 \xi_1 \xi_2^2
\xi_3-2 \xi_2^3 \xi_3 \\
&-24 \xi_3^2+10 \i \xi_1 \xi_3^2+\xi_1^2 \xi_3^2-10 \i \xi_2 \xi_3^2-2 \xi_1 \xi_2
\xi_3^2+\xi_2^2 \xi_3^2 )
\times \frac{ G_6(\xi^{-}_1,\xi^{+}_1,\xi^{-}_2,\xi^{+}_2,\xi^{-}_3,\xi^{+}_3)}{9 (\xi_1-\xi_2)
(-2 \i+\xi_1-\xi_2) (\xi_2-\xi_3) (2 \i+\xi_2-\xi_3)}\\
&+\frac{2 (4 \i+\xi_2-\xi_3) (-6 \i+\xi_2-\xi_3) }{3
(-2 \i+\xi_2-\xi_3) (2 \i+\xi_2-\xi_3)}G_6(\xi^{-}_1,\xi^{+}_1,\xi^{-}_2,\xi^{-}_3,\xi^{+}_2,\xi^{+}_3)
+\frac{(2 \i+\xi_2-\xi_3) (4 \i+\xi_2-\xi_3) }{3
(\xi_2-\xi_3) (-2 \i+\xi_2-\xi_3)}G_6(\xi^{-}_1,\xi^{+}_1,\xi^{-}_3,\xi^{+}_3,\xi^{-}_2,\xi^{+}_2)\\
&+\frac{2 (-4 \i+\xi_1-\xi_2) (6 \i+\xi_1-\xi_2) }{3
(-2 \i+\xi_1-\xi_2) (2 \i+\xi_1-\xi_2)}G_6(\xi^{+}_1,\xi^{+}_2,\xi^{-}_1,\xi^{-}_2,\xi^{-}_3,\xi^{+}_3)
+\frac{(-2 \i+\xi_1-\xi_2) (-4 \i+\xi_1-\xi_2)}{3
(\xi_1-\xi_2) (2 \i+\xi_1-\xi_2)} G_6(\xi^{-}_2,\xi^{+}_2,\xi^{-}_1,\xi^{+}_1,\xi^{-}_3,\xi^{+}_3)\\
&+P_6(\xi^{-}_1,\xi^{+}_1,\xi^{-}_2,\xi^{+}_2,\xi^{-}_3,\xi^{+}_3) \\
%
%
 %
 \\
 &\wtrho{4}(\xi_1,\xi_2,\xi_3)=\frac{1}{27} {\cal N}(\xi_1,\xi_2,\xi_3)
 +\frac{14}{9} G_5(-\xi^{-}_1, -\xi^{-}_2, -\xi^{+}_2,-\xi^{-}_3, -\xi^{+}_3)
 -2 F_6(-\xi^{-}_1, -\xi^{+}_1,-\xi^{-}_2, -\xi^{+}_2,-\xi^{+}_3,-\xi^{-}_3) \\
 &+\frac{4 \left(-2 \i \xi_1+4 \i \xi_2+\xi_1 \xi_2-\xi_2^2-2 \i \xi_3-\xi_1 \xi_3+\xi_2 \xi_3\right)}{(\xi_1-\xi_2)
(\xi_2-\xi_3)}
F_6(\xi^{+}_1,\xi^{-}_1,\xi^{-}_2,\xi^{+}_2,\xi^{-}_3,\xi^{+}_3)\\
&+\frac{2 (4 \i+\xi_2-\xi_3) }{\xi_2-\xi_3}F_6(\xi^{+}_1,\xi^{-}_1,\xi^{-}_2,\xi^{-}_3,\xi^{+}_3,\xi^{+}_2)
+\frac{2
(-4 \i+\xi_1-\xi_2)}{\xi_1-\xi_2} F_6(\xi^{+}_2,\xi^{-}_1,\xi^{+}_1,\xi^{-}_2,\xi^{-}_3,\xi^{+}_3)\\
&+\frac{14
(2 \i+\xi_1-\xi_2) G_6(-\xi^{+}_1,-\xi^{-}_2, -\xi^{-}_1, -\xi^{+}_2,-\xi^{-}_3, -\xi^{+}_3)}{9 (\xi_1-\xi_2)}
+\frac{2 (-4
i+\xi_2-\xi_3) G_6(-\xi^{-}_1, -\xi^{+}_1,-\xi^{+}_2,-\xi^{+}_3,-\xi^{-}_2, -\xi^{-}_3) }{3 (-2 \i+\xi_2-\xi_3)}\\
&-2(56 \i-28 \xi_1-8 \i \xi_1^2-4 \i \xi_1 \xi_2+\xi_1^2 \xi_2+12 \i \xi_2^2-2 \xi_1 \xi_2^2+\xi_2^3+28 \xi_3+20
i \xi_1 \xi_3-\xi_1^2 \xi_3\\
&-20 \i \xi_2 \xi_3+2 \xi_1 \xi_2 \xi_3-\xi_2^2 \xi_3)
\times \frac{G_6(-\xi^{-}_1, -\xi^{+}_1,-\xi^{-}_2, -\xi^{+}_2,-\xi^{-}_3, -\xi^{+}_3)}{9
(\xi_1-\xi_2) (2 \i+\xi_1-\xi_2) (-2 \i+\xi_2-\xi_3)} \\
&-(48 \i \xi_1-24 \xi_1^2-96 \i \xi_2+4 \xi_1 \xi_2-10 \i \xi_1^2 \xi_2-4 \xi_2^2+30
i \xi_1 \xi_2^2+\xi_1^2 \xi_2^2-20 \i \xi_2^3-2 \xi_1 \xi_2^3+\xi_2^4\\
&+48 \i \xi_3+44 \xi_1 \xi_3+10 \i \xi_1^2
\xi_3+4 \xi_2 \xi_3-40 \i \xi_1 \xi_2 \xi_3-2 \xi_1^2 \xi_2 \xi_3+30 \i \xi_2^2 \xi_3+4 \xi_1 \xi_2^2
\xi_3-2 \xi_2^3 \xi_3-24 \xi_3^2\\
&+10 \i \xi_1 \xi_3^2+\xi_1^2 \xi_3^2-10 \i \xi_2 \xi_3^2-2 \xi_1 \xi_2
\xi_3^2+\xi_2^2 \xi_3^2)
\times \frac{G_6(\xi^{-}_1,\xi^{+}_1,\xi^{-}_2,\xi^{+}_2,\xi^{-}_3,\xi^{+}_3)}{9 (\xi_1-\xi_2)
(-2 \i+\xi_1-\xi_2) (\xi_2-\xi_3) (2 \i+\xi_2-\xi_3)}\\
&-\frac{2 (4 \i+\xi_2-\xi_3) (-6 \i+\xi_2-\xi_3)}{3
(-2 \i+\xi_2-\xi_3) (2 \i+\xi_2-\xi_3)} G_6(\xi^{-}_1,\xi^{+}_1,\xi^{-}_2,\xi^{-}_3,\xi^{+}_2,\xi^{+}_3)
-\frac{4 (4 \i+\xi_1-\xi_2)}{2
i+\xi_1-\xi_2} G_6(-\xi^{-}_1, -\xi^{-}_2, -\xi^{+}_1,-\xi^{+}_2,-\xi^{-}_3, -\xi^{+}_3)\\
&-\frac{(2 \i+\xi_2-\xi_3) (4 \i+\xi_2-\xi_3)}{3
(\xi_2-\xi_3) (-2 \i+\xi_2-\xi_3)} G_6(\xi^{-}_1,\xi^{+}_1,\xi^{-}_3,\xi^{+}_3,\xi^{-}_2,\xi^{+}_2)
-\frac{2 (-4 \i+\xi_1-\xi_2) (6 \i+\xi_1-\xi_2)}{3
(-2 \i+\xi_1-\xi_2) (2 \i+\xi_1-\xi_2)} G_6(\xi^{+}_1,\xi^{+}_2,\xi^{-}_1,\xi^{-}_2,\xi^{-}_3,\xi^{+}_3)\\
&-\frac{(-2 \i+\xi_1-\xi_2) (-4 \i+\xi_1-\xi_2) }{3
(\xi_1-\xi_2) (2 \i+\xi_1-\xi_2)}G_6(\xi^{-}_2,\xi^{+}_2,\xi^{-}_1,\xi^{+}_1,\xi^{-}_3,\xi^{+}_3)
-P_6(\xi^{-}_1,\xi^{+}_1,\xi^{-}_2,\xi^{+}_2,\xi^{-}_3,\xi^{+}_3)\\
\displaybreak[2]   
\\
& \wtrho{6}(\xi_1,\xi_2,\xi_3)=
-\frac{1}{27} {\cal N}(\xi_1,\xi_2,\xi_3)-\frac{14}{9} G_5(-\xi^{-}_1, -\xi^{-}_2, -\xi^{+}_2,-\xi^{-}_3, -\xi^{+}_3)+2 F_6(-\xi^{-}_1, -\xi^{+}_1,-\xi^{-}_2, -i-
\xi_2,-\xi^{+}_3,-\xi^{-}_3) 
\\
&-\frac{2 \left(-4 \i \xi_1+8 \i \xi_2+\xi_1 \xi_2-\xi_2^2-4 \i \xi_3-\xi_1 \xi_3+\xi_2 \xi_3\right)}{(\xi_1-\xi_2)
(\xi_2-\xi_3)}
F_6(\xi^{+}_1,\xi^{-}_1,\xi^{-}_2,\xi^{+}_2,\xi^{-}_3,\xi^{+}_3)\\
&-\frac{2 (4 \i+\xi_2-\xi_3) F_6(\xi^{+}_1,\xi^{-}_1,\xi^{-}_2,\xi^{-}_3,\xi^{+}_3,\xi^{+}_2)}{\xi_2-\xi_3}
-\frac{2
(-4 \i+\xi_1-\xi_2) F_6(\xi^{+}_2,\xi^{-}_1,\xi^{+}_1,\xi^{-}_2,\xi^{-}_3,\xi^{+}_3)}{\xi_1-\xi_2}\\
&-\frac{14
(2 \i+\xi_1-\xi_2) G_6(-\xi^{+}_1,-\xi^{-}_2, -\xi^{-}_1, -\xi^{+}_2,-\xi^{-}_3, -\xi^{+}_3)}{9 (\xi_1-\xi_2)}
-\frac{2 (-4
i+\xi_2-\xi_3) G_6(-\xi^{-}_1, -\xi^{+}_1,-\xi^{+}_2,-\xi^{+}_3,-\xi^{-}_2, -\xi^{-}_3) }{3 (-2 \i+\xi_2-\xi_3)}\\
&+2(56 \i-28 \xi_1-8 \i \xi_1^2-4 \i \xi_1 \xi_2+\xi_1^2 \xi_2+12 \i \xi_2^2-2 \xi_1 \xi_2^2+\xi_2^3+28 \xi_3+20i \xi_1 \xi_3 -\xi_1^2 \xi_3\\
&20 \i \xi_2 \xi_3+2 \xi_1 \xi_2 \xi_3-\xi_2^2 \xi_3)\times
\frac{ G_6(-\xi^{-}_1, -\xi^{+}_1,-\xi^{-}_2, -\xi^{+}_2,-\xi^{-}_3, -\xi^{+}_3)}{9
(\xi_1-\xi_2) (2 \i+\xi_1-\xi_2) (-2 \i+\xi_2-\xi_3)} \\
&
+(48 \i \xi_1-24 \xi_1^2-96 \i \xi_2+4 \xi_1 \xi_2-10 \i \xi_1^2 \xi_2-4 \xi_2^2+30
i \xi_1 \xi_2^2+\xi_1^2 \xi_2^2-20 \i \xi_2^3-2 \xi_1 \xi_2^3+\xi_2^4+48 \i \xi_3\\
&+44 \xi_1 \xi_3+10 \i \xi_1^2
\xi_3+4 \xi_2 \xi_3-40 \i \xi_1 \xi_2 \xi_3-2 \xi_1^2 \xi_2 \xi_3+30 \i \xi_2^2 \xi_3+4 \xi_1 \xi_2^2
\xi_3 -2 \xi_2^3 \xi_3-24 \xi_3^2\\
&+10 \i \xi_1 \xi_3^2+\xi_1^2 \xi_3^2-10 \i \xi_2 \xi_3^2-2 \xi_1 \xi_2
\xi_3^2+\xi_2^2 \xi_3^2)\times
\frac{G_6(\xi^{-}_1,\xi^{+}_1,\xi^{-}_2,\xi^{+}_2,\xi^{-}_3,\xi^{+}_3)}{9 (\xi_1-\xi_2)
(-2 \i+\xi_1-\xi_2) (\xi_2-\xi_3) (2 \i+\xi_2-\xi_3)}\\
&+\frac{4 (4 \i+\xi_1-\xi_2) G_6(-\xi^{-}_1, -\xi^{-}_2, -\xi^{+}_1,-\xi^{+}_2,-\xi^{-}_3, -\xi^{+}_3)}{2
i+\xi_1-\xi_2}
+\frac{2 (4 \i+\xi_2-\xi_3) (-6 \i+\xi_2-\xi_3)}{3
(-2 \i+\xi_2-\xi_3) (2 \i+\xi_2-\xi_3)} G_6(\xi^{-}_1,\xi^{+}_1,\xi^{-}_2,\xi^{-}_3,\xi^{+}_2,\xi^{+}_3)\\
&+\frac{(2 \i+\xi_2-\xi_3) (4 \i+\xi_2-\xi_3)}{3
(\xi_2-\xi_3) (-2 \i+\xi_2-\xi_3)} G_6(\xi^{-}_1,\xi^{+}_1,\xi^{-}_3,\xi^{+}_3,\xi^{-}_2,\xi^{+}_2)
+\frac{2 (-4 \i+\xi_1-\xi_2) (6 \i+\xi_1-\xi_2) }{3
(-2 \i+\xi_1-\xi_2) (2 \i+\xi_1-\xi_2)}G_6(\xi^{+}_1,\xi^{+}_2,\xi^{-}_1,\xi^{-}_2,\xi^{-}_3,\xi^{+}_3)
\\
&+\frac{(-2 \i+\xi_1-\xi_2) (-4 \i+\xi_1-\xi_2) }{3
(\xi_1-\xi_2) (2 \i+\xi_1-\xi_2)}G_6(\xi^{-}_2,\xi^{+}_2,\xi^{-}_1,\xi^{+}_1,\xi^{-}_3,\xi^{+}_3)
+P_6(\xi^{-}_1,\xi^{+}_1,\xi^{-}_2,\xi^{+}_2,\xi^{-}_3,\xi^{+}_3)\\
%
%
\\
 &\wtrho{7}(\xi_1,\xi_2,\xi_3)=
\frac{1}{27} {\cal N}(\xi_1,\xi_2,\xi_3)
-\frac{4}{9} G_5(-\xi^{-}_1, -\xi^{-}_2, -\xi^{+}_2,-\xi^{-}_3, -\xi^{+}_3)\\
&+\frac{2 (-4 \i+\xi_2-\xi_3)}{\xi_2-\xi_3} F_6(\xi^{+}_1,\xi^{-}_1,\xi^{-}_2,\xi^{+}_2,\xi^{-}_3,\xi^{+}_3)
+\frac{2
(4 \i+\xi_2-\xi_3) }{\xi_2-\xi_3}F_6(\xi^{+}_1,\xi^{-}_1,\xi^{-}_2,\xi^{-}_3,\xi^{+}_3,\xi^{+}_2)\\
&-\frac{4
(2 \i+\xi_1-\xi_2)}{9 (\xi_1-\xi_2)} G_6(-\xi^{+}_1,-\xi^{-}_2, -\xi^{-}_1, -\xi^{+}_2,-\xi^{-}_3, -\xi^{+}_3)
+\frac{2 (-4
i+\xi_2-\xi_3)}{3 (-2 \i+\xi_2-\xi_3)} G_6(-\xi^{-}_1, -\xi^{+}_1,-\xi^{+}_2,-\xi^{+}_3,-\xi^{-}_2, -\xi^{-}_3) \\
&-\frac{2(-8-8 \i \xi_1+4 \i \xi_2+\xi_1 \xi_2-\xi_2^2+4 \i \xi_3-\xi_1 \xi_3+\xi_2 \xi_3 ) }{9
(\xi_1-\xi_2) (-2 \i+\xi_2-\xi_3)}
G_6(-\xi^{-}_1, -\xi^{+}_1,-\xi^{-}_2, -\xi^{+}_2,-\xi^{-}_3, -\xi^{+}_3)\\
&-(48 \i \xi_1-24 \xi_1^2+48 \i \xi_2+76 \xi_1 \xi_2-10 \i \xi_1^2
\xi_2-4 \xi_2^2-6 \i \xi_1 \xi_2^2+\xi_1^2 \xi_2^2+16 \i \xi_2^3-2 \xi_1 \xi_2^3+\xi_2^4-96 \i \xi_3\\
&-28\xi_1 \xi_3+10 \i \xi_1^2 \xi_3-68 \xi_2 \xi_3+32 \i \xi_1 \xi_2 \xi_3-2 \xi_1^2 \xi_2 \xi_3-42 \i \xi_2^2
\xi_3+4 \xi_1 \xi_2^2 \xi_3-2 \xi_2^3 \xi_3+48 \xi_3^2\\
&-26 \i \xi_1 \xi_3^2+\xi_1^2 \xi_3^2+26 \i \xi_2
\xi_3^2-2 \xi_1 \xi_2 \xi_3^2+\xi_2^2 \xi_3^2)
\times
\frac{ G_6(\xi^{-}_1,\xi^{+}_1,\xi^{-}_2,\xi^{+}_2,\xi^{-}_3,\xi^{+}_3)}{9
(\xi_1-\xi_2) (-2 \i+\xi_1-\xi_2) (\xi_2-\xi_3) (2 \i+\xi_2-\xi_3)}\\
&-\frac{2 (4 \i+\xi_2-\xi_3) (-6 \i+\xi_2-\xi_3)
}{3 (-2 \i+\xi_2-\xi_3) (2 \i+\xi_2-\xi_3)}
G_6(\xi^{-}_1,\xi^{+}_1,\xi^{-}_2,\xi^{-}_3,\xi^{+}_2,\xi^{+}_3)
-\frac{(2\i+\xi_2-\xi_3) (4 \i+\xi_2
-\xi_3) }{3 (\xi_2-\xi_3)
(-2 \i+\xi_2-\xi_3)}G_6(\xi^{-}_1,\xi^{+}_1,\xi^{-}_3,\xi^{+}_3,\xi^{-}_2,\xi^{+}_2)\\
&+\frac{4 (-4 \i+\xi_1-\xi_2) (6 \i+\xi_1-\xi_2) G_6(\xi^{+}_1,\xi^{+}_2,\xi^{-}_1,\xi^{-}_2,\xi^{-}_3,\xi^{+}_3)}{3
(-2 \i+\xi_1-\xi_2) (2 \i+\xi_1-\xi_2)}
+\frac{2 (-2 \i+\xi_1-\xi_2) (-4 \i+\xi_1-\xi_2)}{3
(\xi_1-\xi_2) (2 \i+\xi_1-\xi_2) }
 G_6(\xi^{-}_2,\xi^{+}_2,\xi^{-}_1,\xi^{+}_1,\xi^{-}_3,\xi^{+}_3)\\
& -P_6(\xi^{-}_1,\xi^{+}_1,\xi^{-}_2,\xi^{+}_2,\xi^{-}_3,\xi^{+}_3)\\
\displaybreak[2]   
%
%
\\
&\wtrho{9}(\xi_1,\xi_2,\xi_3)=
\frac{1}{27} {\cal N}(\xi_1,\xi_2,\xi_3)
-\frac{4}{9} G_5(-\xi^{-}_1, -\xi^{-}_2, -\xi^{+}_2,-\xi^{-}_3, -\xi^{+}_3)
-2 F_6(-\xi^{-}_1, -\xi^{+}_1,-\xi^{-}_2, -\xi^{+}_2,-\xi^{+}_3,-\xi^{-}_3) \\
&+\frac{2 \left(-4 \i \xi_1+8 \i \xi_2+\xi_1 \xi_2-\xi_2^2-4 \i \xi_3-\xi_1 \xi_3+\xi_2 \xi_3\right)}{(\xi_1-\xi_2)
(\xi_2-\xi_3)}
F_6(\xi^{+}_1,\xi^{-}_1,\xi^{-}_2,\xi^{+}_2,\xi^{-}_3,\xi^{+}_3)\\
&+\frac{2 (4 \i+\xi_2-\xi_3)}{\xi_2-\xi_3} F_6(\xi^{+}_1,\xi^{-}_1,\xi^{-}_2,\xi^{-}_3,\xi^{+}_3,\xi^{+}_2)
+\frac{2
(-4 \i+\xi_1-\xi_2)}{\xi_1-\xi_2} F_6(\xi^{+}_2,\xi^{-}_1,\xi^{+}_1,\xi^{-}_2,\xi^{-}_3,\xi^{+}_3)\\
&-\frac{4
(2 \i+\xi_1-\xi_2)}{9 (\xi_1-\xi_2)} G_6(-\xi^{+}_1,-\xi^{-}_2, -\xi^{-}_1, -\xi^{+}_2,-\xi^{-}_3, -\xi^{+}_3)
+\frac{2 (-4
i+\xi_2-\xi_3) }{3 (-2 \i+\xi_2-\xi_3)}G_6(-\xi^{-}_1, -\xi^{+}_1,-\xi^{+}_2,-\xi^{+}_3,-\xi^{-}_2, -\xi^{-}_3) \\
&+(32\i-52 \xi_1-2 \i \xi_1^2+36 \xi_2-10 \i \xi_1 \xi_2+7 \xi_1^2 \xi_2+12 \i \xi_2^2-14 \xi_1 \xi_2^2+7 \xi_2^3+16
\xi_3+14 \i \xi_1 \xi_3-7 \xi_1^2 \xi_3\\
&-14 \i \xi_2 \xi_3+14 \xi_1 \xi_2 \xi_3-7 \xi_2^2 \xi_3)\times
\frac{G_6(-\xi^{-}_1, -\xi^{+}_1,-\xi^{-}_2, -\xi^{+}_2,-\xi^{-}_3, -\xi^{+}_3)}{9 (\xi_1-\xi_2) (2 \i+\xi_1-\xi_2)
(-2 \i+\xi_2-\xi_3)}\\
&-(48 \i \xi_1-24 \xi_1^2-96 \i \xi_2+4 \xi_1 \xi_2-10 \i \xi_1^2 \xi_2-4 \xi_2^2+30
i \xi_1 \xi_2^2+\xi_1^2 \xi_2^2-20 \i \xi_2^3-2 \xi_1 \xi_2^3+\xi_2^4\\
&+48 \i \xi_3+44 \xi_1 \xi_3+10 \i \xi_1^2
\xi_3+4 \xi_2 \xi_3-40 \i \xi_1 \xi_2 \xi_3-2 \xi_1^2 \xi_2 \xi_3+30 \i \xi_2^2 \xi_3+4 \xi_1 \xi_2^2
\xi_3-2 \xi_2^3 \xi_3-24 \xi_3^2\\
&+10 \i \xi_1 \xi_3^2+\xi_1^2 \xi_3^2-10 \i \xi_2 \xi_3^2-2 \xi_1 \xi_2
\xi_3^2+\xi_2^2 \xi_3^2)\times
\frac{G_6(\xi^{-}_1,\xi^{+}_1,\xi^{-}_2,\xi^{+}_2,\xi^{-}_3,\xi^{+}_3)}{9 (\xi_1-\xi_2)
(-2 \i+\xi_1-\xi_2) (\xi_2-\xi_3) (2 \i+\xi_2-\xi_3)}\\
&+\frac{2 (4 \i+\xi_1-\xi_2) }{2
i+\xi_1-\xi_2}G_6(-\xi^{-}_1, -\xi^{-}_2, -\xi^{+}_1,-\xi^{+}_2,-\xi^{-}_3, -\xi^{+}_3)
-\frac{2 (4 \i+\xi_2-\xi_3) (-6 \i+\xi_2-\xi_3)}{3
(-2 \i+\xi_2-\xi_3) (2 \i+\xi_2-\xi_3)} G_6(\xi^{-}_1,\xi^{+}_1,\xi^{-}_2,\xi^{-}_3,\xi^{+}_2,\xi^{+}_3)\\
&-\frac{(2 \i+\xi_2-\xi_3) (4 \i+\xi_2-\xi_3) }{3
(\xi_2-\xi_3) (-2 \i+\xi_2-\xi_3)}
G_6(\xi^{-}_1,\xi^{+}_1,\xi^{-}_3,\xi^{+}_3,\xi^{-}_2,\xi^{+}_2)
-\frac{2 (-4 \i+\xi_1-\xi_2) (6 \i+\xi_1-\xi_2) }{3
(-2 \i+\xi_1-\xi_2) (2 \i+\xi_1-\xi_2)}
G_6(\xi^{+}_1,\xi^{+}_2,\xi^{-}_1,\xi^{-}_2,\xi^{-}_3,\xi^{+}_3)\\
&-\frac{(-2 \i+\xi_1-\xi_2) (-4 \i+\xi_1-\xi_2) }{3
(\xi_1-\xi_2) (2 \i+\xi_1-\xi_2)}
G_6(\xi^{-}_2,\xi^{+}_2,\xi^{-}_1,\xi^{+}_1,\xi^{-}_3,\xi^{+}_3)
-P_6(\xi^{-}_1,\xi^{+}_1,\xi^{-}_2,\xi^{+}_2,\xi^{-}_3,\xi^{+}_3) \\
\displaybreak[2]   
%
%
\\
&\wtrho{10}(\xi_1,\xi_2,\xi_3)=
\frac{1}{27} {\cal N}(\xi_1,\xi_2,\xi_3)-\frac{4}{9} G_5(-\xi^{-}_1, -\xi^{-}_2, -\xi^{+}_2,-\xi^{-}_3, -\xi^{+}_3)
\\
&+\frac{2 \left(-4
i \xi_1+8 \i \xi_2+\xi_1 \xi_2-\xi_2^2-4 \i \xi_3-\xi_1 \xi_3+\xi_2 \xi_3\right)}{(\xi_1-\xi_2) (\xi_2-\xi_3)} F_6(\xi^{+}_1,\xi^{-}_1,\xi^{-}_2,\xi^{+}_2,\xi^{-}_3,\xi^{+}_3)\\
&+\frac{2
(4 \i+\xi_2-\xi_3) }{\xi_2-\xi_3}F_6(\xi^{+}_1,\xi^{-}_1,\xi^{-}_2,\xi^{-}_3,\xi^{+}_3,\xi^{+}_2)
+\frac{2 (-4 \i+\xi_1-\xi_2)}{\xi_1-\xi_2}
F_6(\xi^{+}_2,\xi^{-}_1,\xi^{+}_1,\xi^{-}_2,\xi^{-}_3,\xi^{+}_3)\\
&-\frac{4
(2 \i+\xi_1-\xi_2) }{9 (\xi_1-\xi_2)}
G_6(-\xi^{+}_1,-\xi^{-}_2, -\xi^{-}_1, -\xi^{+}_2,-\xi^{-}_3, -\xi^{+}_3)
-\frac{4 (-4
i+\xi_2-\xi_3) }{3 (-2 \i+\xi_2-\xi_3)}G_6(-\xi^{-}_1, -\xi^{+}_1,-\xi^{+}_2,-\xi^{+}_3,-\xi^{-}_2, -\xi^{-}_3) \\
&-\frac{2
\left(-8+10 \i \xi_1-14 \i \xi_2+\xi_1 \xi_2-\xi_2^2+4 \i \xi_3-\xi_1 \xi_3+\xi_2 \xi_3\right) }{ 9
(\xi_1-\xi_2) (-2 \i+\xi_2-\xi_3) }
G_6(-\xi^{-}_1, -\xi^{+}_1,-\xi^{-}_2, -\xi^{+}_2,-\xi^{-}_3, -\xi^{+}_3)\\
&
-(48 \i \xi_1-24 \xi_1^2-96 \i \xi_2+4 \xi_1 \xi_2-10 \i \xi_1^2
\xi_2-4 \xi_2^2+30 \i \xi_1 \xi_2^2+\xi_1^2 \xi_2^2-20 \i \xi_2^3-2 \xi_1 \xi_2^3+\xi_2^4\\
&
+48 \i \xi_3+44
\xi_1 \xi_3+10 \i \xi_1^2 \xi_3+4 \xi_2 \xi_3-40 \i \xi_1 \xi_2 \xi_3-2 \xi_1^2 \xi_2 \xi_3+30 \i \xi_2^2
\xi_3+4 \xi_1 \xi_2^2 \xi_3-2 \xi_2^3 \xi_3-24 \xi_3^2\\
&
+10 \i \xi_1 \xi_3^2+\xi_1^2 \xi_3^2-10 \i \xi_2
\xi_3^2-2 \xi_1 \xi_2 \xi_3^2+\xi_2^2 \xi_3^2)
\times \frac{G_6(\xi^{-}_1,\xi^{+}_1,\xi^{-}_2,\xi^{+}_2,\xi^{-}_3,\xi^{+}_3)}{9
(\xi_1-\xi_2) (-2 \i+\xi_1-\xi_2) (\xi_2-\xi_3) (2 \i+\xi_2-\xi_3)}\\
&-\frac{2 (4 \i+\xi_2-\xi_3) (-6 \i+\xi_2-\xi_3)
}{3 (-2 \i+\xi_2-\xi_3) (2 \i+\xi_2-\xi_3)}G_6(\xi^{-}_1,\xi^{+}_1,\xi^{-}_2,\xi^{-}_3,\xi^{+}_2,\xi^{+}_3)
-\frac{(2
i+\xi_2-\xi_3) (4 \i+\xi_2-\xi_3) }{3 (\xi_2-\xi_3)
(-2 \i+\xi_2-\xi_3)}G_6(\xi^{-}_1,\xi^{+}_1,\xi^{-}_3,\xi^{+}_3,\xi^{-}_2,\xi^{+}_2)\\
&-\frac{2 (-4 \i+\xi_1-\xi_2) (6 \i+\xi_1-\xi_2) }{3
(-2 \i+\xi_1-\xi_2) (2 \i+\xi_1-\xi_2)}G_6(\xi^{+}_1,\xi^{+}_2,\xi^{-}_1,\xi^{-}_2,\xi^{-}_3,\xi^{+}_3)
-\frac{(-2 \i+\xi_1-\xi_2) (-4 \i+\xi_1-\xi_2) }{3
(\xi_1-\xi_2) (2 \i+\xi_1-\xi_2)}G_6(\xi^{-}_2,\xi^{+}_2,\xi^{-}_1,\xi^{+}_1,\xi^{-}_3,\xi^{+}_3)\\
&-P_6(\xi^{-}_1,\xi^{+}_1,\xi^{-}_2,\xi^{+}_2,\xi^{-}_3,\xi^{+}_3)\\
\displaybreak[2]   
\\
&\wtrho{12}(\xi_1,\xi_2,\xi_3)=
-\frac{1}{27} {\cal N}(\xi_1,\xi_2,\xi_3)
+\frac{4}{9} G_5(-\xi^{-}_1, -\xi^{-}_2, -\xi^{+}_2,-\xi^{-}_3, -\xi^{+}_3)
+2 F_6(-\xi^{-}_1, -\xi^{+}_1,-\xi^{-}_2, -\xi^{+}_2,-\xi^{+}_3,-\xi^{-}_3) 
\\
&-\frac{4 \left(-2 \i \xi_1+4 \i \xi_2+\xi_1 \xi_2-\xi_2^2-2 \i \xi_3-\xi_1 \xi_3+\xi_2 \xi_3\right)}{(\xi_1-\xi_2)
(\xi_2-\xi_3)}
F_6(\xi^{+}_1,\xi^{-}_1,\xi^{-}_2,\xi^{+}_2,\xi^{-}_3,\xi^{+}_3)\\
&-\frac{2 (4 \i+\xi_2-\xi_3)}{\xi_2-\xi_3} F_6(\xi^{+}_1,\xi^{-}_1,\xi^{-}_2,\xi^{-}_3,\xi^{+}_3,\xi^{+}_2)
-\frac{2
(-4 \i+\xi_1-\xi_2)}{\xi_1-\xi_2} F_6(\xi^{+}_2,\xi^{-}_1,\xi^{+}_1,\xi^{-}_2,\xi^{-}_3,\xi^{+}_3)\\
&+\frac{4
(2 \i+\xi_1-\xi_2)}{9 (\xi_1-\xi_2)} 
 G_6(-\xi^{+}_1,-\xi^{-}_2, -\xi^{-}_1, -\xi^{+}_2,-\xi^{-}_3, -\xi^{+}_3)\\
&-(32 \i-52 \xi_1-2 \i \xi_1^2+36 \xi_2-10 \i \xi_1 \xi_2+7 \xi_1^2 \xi_2+12 \i \xi_2^2-14 \xi_1 \xi_2^2+7 \xi_2^3+16
\xi_3+14 \i \xi_1 \xi_3-7 \xi_1^2 \xi_3\\
&-14 \i \xi_2 \xi_3+14 \xi_1 \xi_2 \xi_3-7 \xi_2^2 \xi_3)
\times \frac{G_6(-\xi^{-}_1, -\xi^{+}_1,-\xi^{-}_2, -\xi^{+}_2,-\xi^{-}_3, -\xi^{+}_3)}{9 (\xi_1-\xi_2) (2 \i+\xi_1-\xi_2)
(-2 \i+\xi_2-\xi_3)}\\
&-\frac{2 (4 \i+\xi_1-\xi_2)}{2
i+\xi_1-\xi_2} G_6(-\xi^{-}_1, -\xi^{-}_2, -\xi^{+}_1,-\xi^{+}_2,-\xi^{-}_3, -\xi^{+}_3)\\
&+ 2 (24 \i \xi_1-12 \xi_1^2-48 \i \xi_2-16 \xi_1 \xi_2-14 \i \xi_1^2 \xi_2+16 \xi_2^2+24
i \xi_1 \xi_2^2+5 \xi_1^2 \xi_2^2-10 \i \xi_2^3-10 \xi_1 \xi_2^3+5 \xi_2^4\\
&+24 \i \xi_3+40 \xi_1 \xi_3+14
i \xi_1^2 \xi_3-16 \xi_2 \xi_3-20 \i \xi_1 \xi_2 \xi_3-10 \xi_1^2 \xi_2 \xi_3+6 \i \xi_2^2 \xi_3+20
\xi_1 \xi_2^2 \xi_3-10 \xi_2^3 \xi_3-12 \xi_3^2\\
&
-4 \i \xi_1 \xi_3^2+5 \xi_1^2 \xi_3^2+4 \i \xi_2 \xi_3^2-10
\xi_1 \xi_2 \xi_3^2+5 \xi_2^2 \xi_3^2)
 \times \frac{G_6(\xi^{-}_1,\xi^{+}_1,\xi^{-}_2,\xi^{+}_2,\xi^{-}_3,\xi^{+}_3)}{9
(\xi_1-\xi_2) (-2 \i+\xi_1-\xi_2) (\xi_2-\xi_3) (2 \i+\xi_2-\xi_3)}\\
&-\frac{2 (-4
i+\xi_2-\xi_3) G_6(-\xi^{-}_1, -\xi^{+}_1,-\xi^{+}_2,-\xi^{+}_3,-\xi^{-}_2, -\xi^{-}_3) }{3 (-2 \i+\xi_2-\xi_3)}
+\frac{8 (-3 \i+\xi_2-\xi_3) (4 \i+\xi_2-\xi_3)
}{3 (-2 \i+\xi_2-\xi_3) (2 \i+\xi_2-\xi_3)}G_6(\xi^{-}_1,\xi^{+}_1,\xi^{-}_2,\xi^{-}_3,\xi^{+}_2,\xi^{+}_3)\\
&+\frac{(2
i+\xi_2-\xi_3) (4 \i+\xi_2-\xi_3) }{3 (\xi_2-\xi_3)
(-2 \i+\xi_2-\xi_3)}G_6(\xi^{-}_1,\xi^{+}_1,\xi^{-}_3,\xi^{+}_3,\xi^{-}_2,\xi^{+}_2)
+\frac{2 (-4 \i+\xi_1-\xi_2) (6 \i+\xi_1-\xi_2)}{3
(-2 \i+\xi_1-\xi_2) (2 \i+\xi_1-\xi_2)} G_6(\xi^{+}_1,\xi^{+}_2,\xi^{-}_1,\xi^{-}_2,\xi^{-}_3,\xi^{+}_3)\\
&+\frac{(-2 \i+\xi_1-\xi_2) (-4 \i+\xi_1-\xi_2) }{3
(\xi_1-\xi_2) (2 \i+\xi_1-\xi_2)}G_6(\xi^{-}_2,\xi^{+}_2,\xi^{-}_1,\xi^{+}_1,\xi^{-}_3,\xi^{+}_3)
+P_6(\xi^{-}_1,\xi^{+}_1,\xi^{-}_2,\xi^{+}_2,\xi^{-}_3,\xi^{+}_3)
\end{align*}
\end{footnotesize}
The rest is easily reproduced by applying the symmetry relations (\ref{rhosymmetry}).
The explicit forms of $F_a, G_a$ and $P_a$ in terms of $\omega(\xi_i,\xi_j)$
are too lengthly to be reproduced here.\\

\bibliographystyle{ourbook} \bibliography{hub}

\begin{thebibliography}{10}

\bibitem{AuKl12}
Britta Aufgebauer and Andreas Kl\"umper: {\it Finite temperature correlation
functions from discrete functional equations},  J. Phys. A:
Math. Theor. 45 (2012) 345203.

\bibitem{Babujian82}
H.~M. Babujian, \emph{Exact solution of the one-dimensional isotropic
  {H}eisenberg chain witgh arbitrary spins {S}}, Phys. Lett. A \textbf{90}
  (1982) 479.

\bibitem{BoGo09}
H.~Boos and F.~G\"ohmann, \emph{On the physical part of the factorized
  correlation functions of the {XXZ} chain}, J. Phys. A \textbf{42} (2009)
  315001.

\bibitem{BGKS06}
H.~Boos, F.~G\"ohmann, A.~Kl\"umper and J.~Suzuki, \emph{Factorization of
  multiple integrals representing the density matrix of a finite segment of the
  {H}eisenberg spin chain}, J. Stat. Mech.  (2006) P04001.

\bibitem{BGKS07}
---, \emph{Factorization of the finite temperature correlation functions of the
  {XXZ} chain in a magnetic field}, J. Phys. A \textbf{40} (2007) 10699.

\bibitem{BJMST06b}
H.~Boos, M.~Jimbo, T.~Miwa, F.~Smirnov and Y.~Takeyama, \emph{Hidden
  {G}rassmann structure in the {XXZ} model}, Comm. Math. Phys. \textbf{272}
  (2007) 263.

\bibitem{BJMST08a}
---, \emph{Hidden {G}rassmann structure in the {XXZ} model {II}: creation
  operators}, Comm. Math. Phys. \textbf{286} (2009) 875.

\bibitem{BoKo01}
H.~E. Boos and V.~E. Korepin, \emph{Quantum spin chains and {R}iemann zeta
  function with odd arguments}, J. Phys. A \textbf{34} (2001) 5311.

\bibitem{BoWe94}
A.~H. Bougourzi and R.~A. Weston, \emph{N-point correlation functions of the
  spin-1 {XXZ} model}, Nucl. Phys. B \textbf{417} (1994) 439.

\bibitem{DGHK07}
J.~Damerau, F.~G\"ohmann, N.~P. Hasenclever and A.~Kl\"umper, \emph{Density
  matrices for finite segments of {H}eisenberg chains of arbitrary length}, J.
  Phys. A \textbf{40} (2007) 4439.

\bibitem{DeMa10}
T.~Deguchi and C.~Matsui, \emph{Correlation functions of the integrable
  higher-spin {XXX} and {XXZ} spin chains through the fusion method}, Nucl.
  Phys. B \textbf{831} (2010) 359.

\bibitem{thebook}
F.~H.~L. Essler, H.~Frahm, F.~G\"ohmann, A.~Kl\"umper and V.~E. Korepin,
  \emph{The {O}ne-{D}imensional {H}ubbard {M}odel} (Cambridge University Press,
  2005).


\bibitem{GHS05}
F.~G\"ohmann, N.~P. Hasenclever and A.~Seel, \emph{The finite temperature
  density matrix and two-point correlations in the antiferromagnetic {XXZ}
  chain}, J. Stat. Mech.  (2005) P10015.

\bibitem{GKS04a}
F.~G\"ohmann, A.~Kl\"umper and A.~Seel, \emph{Integral representations for
  correlation functions of the {XXZ} chain at finite temperature}, J. Phys. A
  \textbf{37} (2004) 7625.

\bibitem{GKS05}
---, \emph{Integral representation of the density matrix of the {XXZ} chain at
  finite temperature}, J. Phys. A \textbf{38} (2005) 1833.

\bibitem{GSS10}
F.~G\"ohmann, A.~Seel and J.~Suzuki, \emph{Correlation functions of the
integrable isotropic spin-1 chain at finite temperature}, J. Stat. Mech. (2010)
P11011

\bibitem{GoSu10}
F.~G\"ohmann and J.~Suzuki, \emph{Quantum spin chains at finite temperature},
in ``New Trends in Quantum Integrable Systems'' World Scientific, Singapore
(2010) 81-100.

\bibitem{Idzumi94}
M.~Idzumi, \emph{Level-2 irreducible representations of {U}(q)(sl(2)), vertex
  operators, and their correlations}, Int. J. Mod. Phys. A \textbf{9} (1994)
  4449.

\bibitem{JMMN92}
M.~Jimbo, K.~Miki, T.~Miwa and A.~Nakayashiki, \emph{Correlation functions of
  the {XXZ} model for {$\Delta < - 1$}}, Phys. Lett. A \textbf{168} (1992) 256.

\bibitem{JiMi96}
M.~Jimbo and T.~Miwa, \emph{Quantum {KZ} equation with $|q| = 1$ and
  correlation functions of the {XXZ} model in the gapless regime}, J. Phys. A
  \textbf{29} (1996) 2923.

\bibitem{JMS08}
M.~Jimbo, T.~Miwa and F.~Smirnov, \emph{Hidden {G}rassmann structure in the
  {XXZ} model {III}: introducing {M}atsubara direction}, J. Phys. A \textbf{42}
  (2009) 304018.

\bibitem{JMS09pp}
---, \emph{On one-point functions of descendants in {S}ine-{G}ordon model},
in ``New Trends in Quantum Integrable Systems'' World Scientific, Singapore (2010) 117-138.

\bibitem{JKS98a} G.~J\"uttner, A. Kl\"umper and J. Suzuki, \emph{ From fusion
  hierarchy to excited state TBA}, \rm Nucl. Phys. B \textbf{512} (1998) 581

\bibitem{Kitanine01}
N.~Kitanine, \emph{Correlation functions of the higher spin {XXX} chains}, J.
  Phys. A \textbf{34} (2001) 8151.

\bibitem{KKMST07}
N.~Kitanine, K.~Kozlowski, J.~M. Maillet, N.~A. Slavnov and V.~Terras, \emph{On
  correlation functions of integrable models associated with the six-vertex
  ${R}$-matrix}, J. Stat. Mech.  (2007) P01022.

\bibitem{KKMST09}
---, \emph{Algebraic {B}ethe ansatz approach to the asymptotic behavior of
  correlation functions}, J. Stat. Mech.  (2009) P04003.

\bibitem{KMT99b}
N.~Kitanine, J.~M. Maillet and V.~Terras, \emph{Correlation functions of the
  {XXZ} {H}eisenberg spin-$\frac{1}{2}$ chain in a magnetic field}, Nucl. Phys.
  B \textbf{567} (2000) 554.

\bibitem{KP92}
A. Kl\"umper and P. A. Pearce, \emph{Conformal weights of RSOS lattice 
models and their fusion hierarchy}, Physica A \textbf{183}, (1992) 304-350 

\bibitem{Kluemper93}
A.~Kl\"umper, \emph{Thermodynamics of the anisotropic spin-1/2 {H}eisenberg
  chain and related quantum chains}, Z. Phys. B \textbf{91} (1993) 507.

\bibitem{KRS81}
P.~P. Kulish, {N.\ Yu.\ Reshetikhin} and E.~K. Sklyanin, \emph{Yang-{B}axter
  equation and representation theory: {I}}, Lett. Math. Phys. \textbf{5} (1981)
  393.

\bibitem{KuSk82b}
P.~P. Kulish and E.~K. Sklyanin, \emph{Quantum spectral transform method --
  recent developments}, in \emph{Lecture Notes in Physics 151}, 61--119
  (Springer Verlag, Berlin, 1982).

\bibitem{SBGK07a}
A.~Seel, T.~Bhattacharyya, F.~G\"ohmann and A.~Kl\"umper, \emph{A note on the
  spin-1/2 {XXZ} chain concerning its relation to the {B}ose gas}, J. Stat.
  Mech.  (2007) P08030.

\bibitem{SGK07}
A.~Seel, F.~G\"ohmann and A.~Kl\"umper, \emph{From multiple integrals to
  {F}redholm determinants}, Prog. Theor. Phys. Suppl. \textbf{176} (2008) 375.

\bibitem{Suzuki99}
J.~Suzuki, \emph{Spinons in magnetic chains of arbitrary spins at finite
  temperatures}, J. Phys. A \textbf{32} (1999) 2341.

\bibitem{Suzuki85}
M.~Suzuki, \emph{Transfer-matrix method and {Monte Carlo} simulation in quantum
  spin systems}, Phys. Rev. B \textbf{31} (1985) 2957.

\bibitem{SAW}
J.~Suzuki, Y.~Akutsu and M.~Wadati,
\emph{A New approach to Quantum Spin Chains at Finite Temperature},
J. Phys. Soc. Jpn 59 (1990) 2667-2680.

\bibitem{SuIn87}
M.~Suzuki and M.~Inoue, \emph{The {ST}-transformation approach to analytic
  solutions of quantum systems. {I}.\ {G}eneral formulations and basic limit
  theorems}, Prog. Theor. Phys. \textbf{78} (1987) 787.

\bibitem{Takhtajan82}
L.~A. Takhtajan, \emph{The picture of low-lying excitations in the isotropic
  {H}eisenberg chain of arbitrary spins}, Phys. Lett. A \textbf{87} (1982) 479.

\bibitem{ZaFa80}
A.~B. Zamolodchikov and A.~V. Fateev, \emph{A model factorized ${S}$-matrix and
  an integrable spin-$1$ {H}eisenberg chain}, Yad. Fiz. \textbf{32} (1980) 581.

\end{thebibliography}

\end{document}